\documentclass[10pt,twocolumn,superscriptaddress]{revtex4-2}

\usepackage{graphicx}
\usepackage{amsmath}

\usepackage{booktabs}

\usepackage{color}
\definecolor{propblue}{rgb}{0.05,0.20,0.55}

\usepackage[normalem]{ulem}
\usepackage{siunitx}
\usepackage{makecell}
\usepackage{chemformula}

\newif\ifSI
\SItrue
\ifSI\else
  \usepackage{xr-hyper}
  \begingroup\def\celsius{}\def\degree{}\def\metre{}\def\micro{}
  \endgroup
\fi
\usepackage[colorlinks=true,linkcolor=blue,citecolor=blue,urlcolor=blue]{hyperref}
\newcommand{\secref}[1]{Section~\hyperref[#1]{\nameref{#1}}}

\begin{document}

\title{Experimental Evaluation of High-Throughput Structure Predictions}

\author{Gabriele D. Di Gianluca}
\affiliation{Department of Physics, University of Florida}
\author{James Grossmann}
\affiliation{Department of Physics, University of Florida}
\author{Jackson Niedel}
\affiliation{Department of Physics, University of Florida}
\author{Juan Esquivel}
\affiliation{Department of Physics, University of Florida}
\author{Molly Crossman}
\affiliation{Department of Physics, University of Florida}
\author{Tyler Miniati}
\affiliation{Department of Physics, University of Florida}
\author{Melchior Jean}
\affiliation{Department of Physics, University of Florida}
\author{Zhongwei Li}
\affiliation{Department of Physics, University of Florida}
\author{Pawan Prakash}
\affiliation{Department of Physics, University of Florida}
\affiliation{Quantum Theory Project, University of Florida}
\author{Andrew Valentini}
\affiliation{Department of Physics, University of Florida}
\author{Jung Soo Kim}
\affiliation{Department of Physics, University of Florida}
\author{Benjamin Geisler}
\affiliation{Department of Physics, University of Florida}
\affiliation{Department of Materials Science and Engineering, University of Florida}
\author{Richard G. Hennig}
\affiliation{Department of Physics, University of Florida}
\affiliation{Quantum Theory Project, University of Florida}
\affiliation{Department of Materials Science and Engineering, University of Florida}
\author{Peter J. Hirschfeld}
\affiliation{Department of Physics, University of Florida}
\author{Gregory R. Stewart}
\affiliation{Department of Physics, University of Florida}
\author{James J. Hamlin}
\email{jhamlin@ufl.edu}
\affiliation{Department of Physics, University of Florida}

\begin{abstract}
High throughput computational databases such as Materials Project, the Open Quantum Materials Database, and Alexandria now each list on the order of $10^5$ predicted stable compounds.
However, the overall reliability of these predictions in terms of actual experimental synthesizability remains poorly understood.
In this work we attempt to synthesize, via arc melting, 34 different predicted ternary compounds.
The majority of these materials did not form in the predicted structure, instead resulting in disordered variants or multiphase mixtures.
Four formed ordered structures not reported before, one of them at a composition reported earlier without a structure.
We discuss both the failures and the successes, and identify simple filtering criteria that can help researchers avoid hypothetical materials that are unlikely to form.
\end{abstract}

\maketitle

\section{Introduction}
The discovery of new, functional materials such as high-$T_\mathrm{c}$ superconductors, improved thermoelectrics, more efficient battery materials, materials for spintronics, etc, remains a central challenge in materials science~\cite{Curtarolo2013, Oganov2018}.
Historically, the synthesis of new materials has been driven by a combination of experimental expertise, trial and error, and serendipity~\cite{Garca2008}.
Despite the long history of successes that has produced nearly all of the functional materials in use today, the empirical method of materials discovery is costly and slow~\cite{Curtarolo2013, Shahzad2024}.

Computational methods have emerged as a promising alternative, predicting the stability of new materials and thereby guiding experimental discovery~\cite{Curtarolo2013,Oganov2018}.
In particular, advances in available computational resources have enabled high-throughput density functional theory (DFT) calculations to systematically determine the formation energies for large numbers of known and hypothetical crystal structures.
This has given rise to material databases that now each list on the order of $10^5$ predicted stable compounds~\cite{Shen2022, Horton2025, Cavignac2026}.
In recent years, generative AI has facilitated the generation of even more new crystal structures, while machine-learning (ML) models have reduced the computational cost of evaluating predicted structures.
Together, these advances have drastically expanded the scale of computational materials discovery~\cite{Merchant2023, Zeni2025, Metni2026}.
As a result, there now exist open materials databases such as Materials Project (MP), Open Quantum Materials Database (OQMD), and Alexandria containing millions of hypothetical structures~\cite{Saal2013, Jain2013, Schmidt2024}. 

Based on the abundance of candidate materials in these open databases, an enormous and wide-ranging effort has been undertaken to apply high-throughput computational screening methods to search for materials with desired properties.
For example, structures taken from large materials databases have been screened for superconductivity~\cite{Choudhary2022,Wines2023,Cerqueira2023,Gibson2026, Gibson2025}, magnetic and topological order~\cite{Zhang2019, Frey2020, Su2022, Vergniory2019, Xu2020, Robredo2025, Choudhary2021}, and altermagnetic spin splitting~\cite{Sufyan2026, Sdequist2024, Wan2025}.
Experimental groups have likewise invested considerable time and effort attempting to synthesize and characterize individual candidates predicted to have promising properties.
While this effort has undoubtedly produced interesting materials, many of the compounds studied had already been reported as experimentally realized and are simply being characterized in a new way~\cite{Park2026, Cheetham2024}.

A DFT calculation predicts an ordered crystal structure, but in the compound that forms, two or more elements can mix on one crystallographic site, producing compositional disorder.
Leeman et al.~\cite{Leeman2024} argue that such a disordered compound is a different material from the predicted compound, and that a claim of a new material should state how it differs from materials already known.
In this paper, we count a prediction as a success only when the predicted ordered structure formed and the compound had not been reported before.

Most of the compounds that Gautier et al.~\cite{Gautier2015} predicted and then attempted meet this standard.
They restricted their search to one chemical family, the 483 ternary ABX compositions with 18 valence electrons per formula unit, 400 of them unreported.
For each ternary system they computed the energies of the reported phases together with hypothetical phases constructed by analogy with reported isovalent compounds.
Of the 16 predicted-stable compounds they attempted by arc melting and annealing, 15 formed in the predicted ordered structure, at compositions that had not been reported in the experimental literature.

Other studies that tested a list of predictions found far fewer successes.
Narayan et al.~\cite{Narayan2016} took a similar per-system approach to sulfides and selenides with no known ternary compounds, and none of the 15 candidates they attempted formed.
In oxides, Montoya et al.~\cite{Montoya2024} found new compounds in only one of six predicted ternary systems, and not the predicted ones, and Walters et al.~\cite{Walters2025} obtained only one of the cation substitutions into \ch{Na4TiP2O9} predicted to lie near the hull.
Gautier et al.~\cite{Gautier2015} and these three groups all computed their own hulls, whereas most screening studies take their candidates from the hull energies stored in public databases.
One possible explanation for the success of Gautier et al.~\cite{Gautier2015} is that their targets came from a chemical family in which chemical intuition about stable compounds is well developed~\cite{Cheetham2024}.

Few experiments have tested, across many ternary systems, how often a never-reported compound listed on or near the database hulls forms.
To our knowledge, the success rate of the largest experimental test of database predictions has been disputed.
Szymanski et al.~\cite{Szymanski2023} reported 41 new compounds from 58 targets that lie on or near the MP hull and were cross-checked against a similar database from Google DeepMind.
Leeman et al.~\cite{Leeman2024} re-examined the diffraction data and concluded that none were new.
In their analysis, about two thirds of the reported compounds were more likely known, disordered versions of the predicted ordered compounds, and most of the rest had been reported before.
Other groups reached similar conclusions~\cite{Yamamoto2024,Widdowson2025}.
In a correction, Szymanski et al.~\cite{Szymanski2023correction} revised the count to 36 of 57.
They clarified that the targets were new to their prediction platform, and that partially disordered versions of a target were counted as successes.
Their definition of success is therefore less strict than the one we use in this study.

Later studies have run into the same difficulties with compositional disorder and with compounds that had already been reported.
Prein et al.~\cite{Prein2025} report the synthesis of 7 of 16 database targets, but they claim that six of the seven had been reported before.
A disordered form of \ch{TaCr2O6} proposed by a generative model~\cite{Zeni2025} had already been synthesized in 1971~\cite{Juelsholt2026}.
With powder X-ray diffraction (XRD), compositional disorder can be judged only by fitting both ordered and disordered models to the data, and a compound missing from the Inorganic Crystal Structure Database (ICSD)~\cite{Zagorac2019} may still have been reported in the literature.

In this work, we address the question of the reliability of current high-throughput structure predictions by attempting the synthesis of 34 candidate materials taken from MP, OQMD, and Alexandria databases.
The chosen materials were reported to lie on or very close to the convex hull, and all are suitable for arc melting.
We chose arc melting because it is inexpensive, fast relative to other synthesis methods such as solid state synthesis which involves repeated re-grinding and firing, and avoids the experimental complexity and trial and error that accompany growth methods like flux growth or chemical vapor transport. 

Three of the compounds we attempted, \ch{GdPt2Ge}, \ch{TiPt5Si} and \ch{ZrPt5Si}, formed an ordered on-hull structure and had not been reported before.
A fourth, \ch{YCoSi3}, crystallized in a structure not reported before, although the composition was reported in 1974~\cite{Bodak1974}.
From the XRD data we also identify several recurring ways in which the predicted structures fail to form.
Our work highlights important limitations in current high-throughput materials discovery and serves as a guide to researchers on how to identify materials with a greater chance of being experimentally realizable.

\section{Results}

\begin{figure*}[p]
    \centering
    \includegraphics[width=\textwidth,height=0.86\textheight,keepaspectratio]{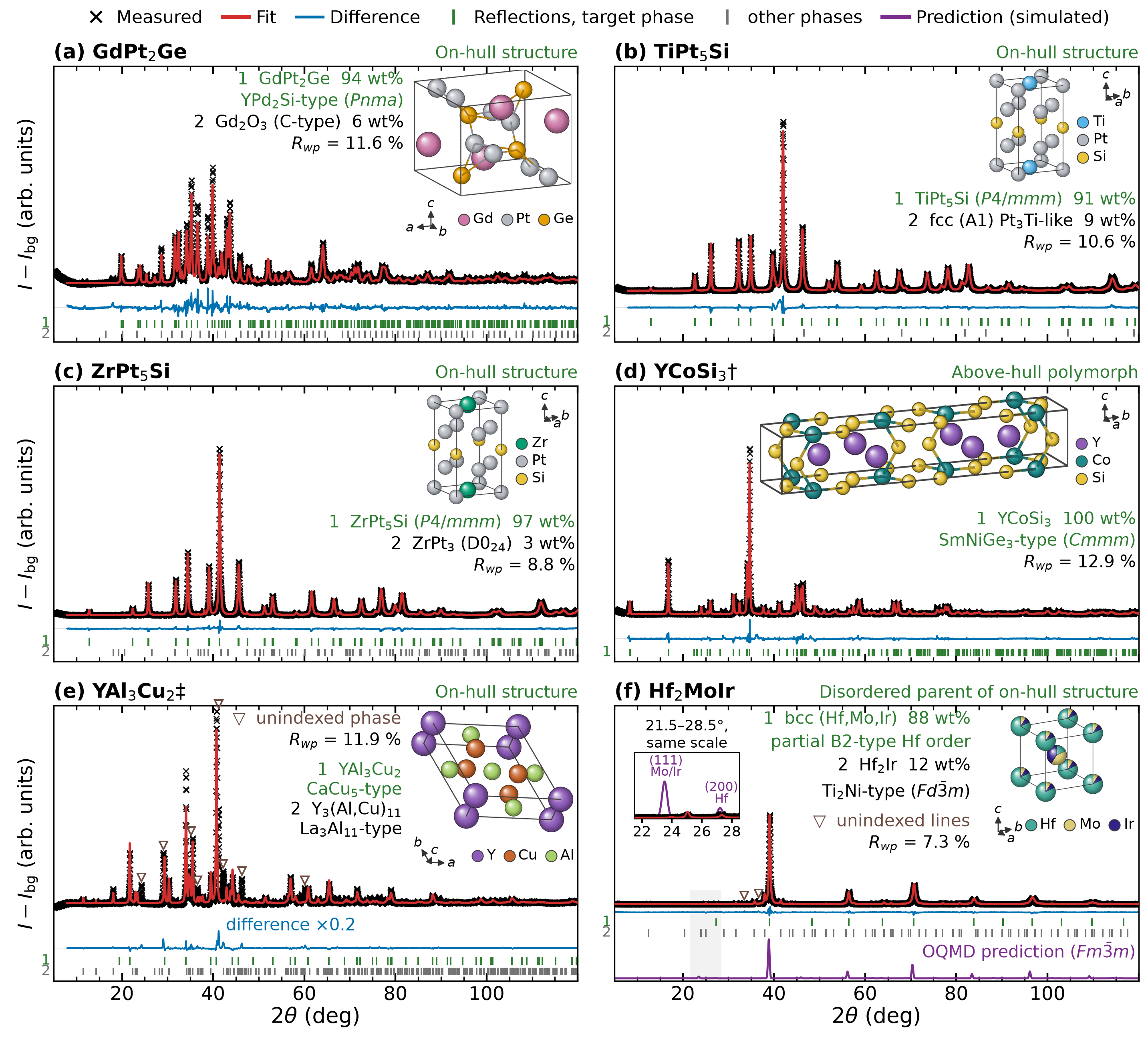}
    \caption{Representative XRD patterns with Rietveld fits. Crosses show the measured pattern minus the fitted background, the red line the fit, the blue line the difference, and the tick rows the Bragg positions of each phase (green for the target phase), with the refined weight fractions. Each drawing shows one unit cell of the phase that formed. Annealed samples were held at 1000\,$^\circ$C for one week and measured on a zero-background slide.
(a) \ch{GdPt2Ge}, annealed. A new compound in the $Pnma$ structure that is the lowest-energy structure at this composition in Alexandria, not the $Cmcm$ structure predicted by OQMD.
(b) \ch{TiPt5Si}, annealed. A new compound in the predicted $P4/mmm$ structure, 91~wt\% of the sample.
(c) \ch{ZrPt5Si}, annealed. A new compound in the predicted $P4/mmm$ structure, 97~wt\% of the sample.
(d) \ch{YCoSi3}$^\dagger$, annealed. A nearly single-phase $Cmmm$ polymorph, 8\,meV/atom above the predicted on-hull $I4mm$ structure.
(e) \ch{YAl3Cu2}$^\ddagger$, annealed. Al and Cu mostly occupy their predicted sites. Lines of an unidentified phase are marked ($\bigtriangledown$), so no weight fractions are given, and the difference curve is scaled by 0.2.
(f) \ch{Hf2MoIr}, annealed. A disordered parent of the predicted structure, bcc with partial Hf order (88~wt\%), with Ti$_2$Ni-type \ch{Hf2Ir} (12~wt\%). The pattern simulated from the OQMD prediction (L2$_1$, $Fm\bar{3}m$, purple) has a Mo/Ir superlattice line at 23.6$^\circ$ that is absent from the data (inset). Triangles mark unindexed lines.
The gray band marks the range shown in the inset.
$^\dagger$Composition reported without a structure. $^\ddagger$Ordered counterpart of a known compound in which Al and Cu share a site.}
    \label{fig:xrd_representative}
\end{figure*}

We arc melted 34 ternary compositions that OQMD, MP or Alexandria lists as stable or close to stable, and characterized each sample by powder XRD and Rietveld refinement (Fig.~\ref{fig:xrd_representative} and Supplementary Section~\ref{sec:S_samples}).
The targets come from three candidate lists (Methods).
List~A contains entries on or near the MP and OQMD hulls in ternary systems with no reported ternary compound, and List~B contains entries on the OQMD hull in systems with exactly one ternary compound.
List~C contains entries on the Alexandria PBEsol hull that the disorder model described below predicts to be ordered.

Following Gautier et al.~\cite{Gautier2015}, we describe the predicted stability of each target by $\Delta H_\mathrm{f}-C(x)$, the energy of its lowest structure relative to the hull of all other compositions in its ternary system (Table~\ref{tab:xrd_outcomes} and Methods).
This is the quantity OQMD lists as the ``stability'' of an entry.
A negative value means that the structure is predicted to be stable by that margin.
In assigning outcomes, we count a target as having formed an on-hull structure if the structure that formed is the lowest-energy structure in at least one database that places it below the hull.
Since the databases do not always agree, a composition can have more than one such structure.
We also give $P_{\text{disorder}}$, the probability that two or more elements share a crystallographic site, which the model of Jakob et al.~\cite{Jakob2025} predicts from the chemical composition alone (Methods).

Table~\ref{tab:xrd_outcomes} sorts the targets into six outcomes, judged from the best-characterized pattern of each (Methods).
Targets that formed at their own composition adopted either an on-hull structure or an above-hull polymorph.
The others formed a disordered parent of the on-hull structure or some other solid solution, separated into other phases (multiphase), or formed phases we could not identify (unresolved).

\begin{table*}[!tp]
\centering
\footnotesize
\caption{Arc-melted targets grouped by XRD outcome category, ordered within a category by decreasing $P_{\mathrm{disorder}}$. The outcome is from the best-characterized scan (after annealing where stated). List: candidate list the target came from (A, B or C; the lists are described in Methods). $\Delta H_{\mathrm{f}}-C(x)$: energy of the lowest entry at the target composition relative to the convex hull $C(x)$ of all other compositions~\cite{Gautier2015} in OQMD, the Materials Project (MP) and Alexandria (PBE), in meV/atom; negative means predicted stable, blank means no entry. $P_{\mathrm{disorder}}$: predicted probability of a chemically disordered structure~\cite{Jakob2025}. A structure is on the hull if it is the lowest-energy structure in at least one database that places it below the hull. $^{*}$: structure previously reported at the target composition; $^{\dagger}$: composition reported without a structure; $^{\ddagger}$: predicted ordered member of a previously reported series whose known member has the same sites partly disordered.}
\label{tab:xrd_outcomes}
\vspace{2pt}
\setlength{\tabcolsep}{4pt}
\renewcommand{\arraystretch}{1.0}
\begin{tabular}{rllrrrrl}
\toprule
 & & & \multicolumn{3}{c}{$\Delta H_{\mathrm{f}}-C(x)$ (meV/atom)} & & \\
\cmidrule(lr){4-6}
\# & Formula & List & OQMD & MP & Alex. & $P_{\mathrm{disorder}}$ & XRD outcome \\
\midrule
\multicolumn{8}{l}{\textbf{On-hull structure}} \\
1 & \ch{ZrPt5Si} & B & -41.8 & -41.8 & -21.5 & 0.20 & On-hull ZrPt$_{5}$Si, 97 wt\% after anneal (ZrPt$_{3}$ 3 wt\%); new compound \\
2 & \ch{YAl3Cu2}$^{\ddagger}$ & C & -10.8 & & -3.0 & 0.18 & \makecell[tl]{Predicted Al/Cu order on the CaCu$_{5}$ lattice (order parameter 0.7 \\ after anneal); La$_{3}$Al$_{11}$-type Y$_{3}$(Al,Cu)$_{11}$ and unidentified lines} \\
3 & \ch{GdPt2Ge} & B & -32.5 & & -25.4 & 0.18 & \makecell[tl]{YPd$_{2}$Si-type $Pnma$ (lowest in Alexandria), 94 wt\% after anneal, small \\ oxide fraction; OQMD $Cmcm$ structure absent; new compound} \\
4 & \ch{TiPt5Si} & B & -16.5 & & -25.6 & 0.18 & On-hull TiPt$_{5}$Si, 91 wt\% after anneal; new compound \\
5 & \ch{YSnCu}$^{*}$ & C & -69.1 & -69.0 & -69.6 & 0.15 & On-hull LiGaGe-type YSnCu formed; previously known compound \\
6 & \ch{ScCrC2}$^{*}$ & C & -10.0 & 26.1 & 26.8 & 0.05 & \makecell[tl]{On-hull $P6_{3}/mmc$ ScCrC$_{2}$, nearly single phase \\ (about 95 wt\%); previously known} \\
\midrule
\multicolumn{8}{l}{\textbf{Above-hull polymorph}} \\
7 & \ch{YCoSi}$^{*}$ & C & -42.9 & 16.9 & -31.1 & 0.13 & \makecell[tl]{Known $Pnma$ YCoSi polymorph (+32 meV) formed, \\ textured; not the on-hull $P4/nmm$ structure} \\
8 & \ch{YCoSi3}$^{\dagger}$ & C & -17.3 & & -14.9 & 0.01 & \makecell[tl]{$Cmmm$ polymorph (+8 meV), nearly single phase after \\ anneal; on-hull $I4mm$ structure absent} \\
\midrule
\multicolumn{8}{l}{\textbf{Disordered parent of on-hull structure}} \\
9 & \ch{Al8FeCo7} & B & -3.4 & 12.1 & & 0.92 & B2 Al(Fe,Co), Al ordered ($\eta$ 0.9); Fe/Co order undetermined \\
10 & \ch{Hf2MoIr} & A & -12.0 & -9.1 & 3.7 & 0.90 & \makecell[tl]{bcc Hf-Mo-Ir after anneal, partial Hf order; \\ Mo/Ir order absent; Ti$_{2}$Ni-type Hf$_{2}$Ir 12 wt\%} \\
11 & \ch{Al4FeCo3} & B & -6.5 & & -3.2 & 0.89 & B2 Al(Fe,Co), Al ordered ($\eta$ 0.9); Fe/Co order undetermined \\
12 & \ch{Sc2PdCu} & A & 4.8 & -1.3 & -0.7 & 0.89 & B2 Sc(Cu,Pd), Sc ordered; Heusler Cu/Pd order absent; cut-face scan \\
13 & \ch{Sc3Pd2Cu} & A & -3.0 & & 4.9 & 0.88 & \makecell[tl]{B2 Sc(Cu,Pd) after anneal, Sc ordered; Cu/Pd layer \\ order of the on-hull structure absent} \\
14 & \ch{ScPd5Au2} & A & 2.8 & & 9.9 & 0.81 & fcc (Sc,Pd,Au) solution; $Pbam$ order absent in range; bulk scan \\
15 & \ch{Al2MnAu6} & B & -0.6 & & 3.2 & 0.62 & C11$_b$ Au$_{2}$(Al,Mn), partial Au order; Mn/Al order undetermined \\
16 & \ch{Sc2Pd5Au} & A & -9.6 & -12.9 & -34.7 & 0.60 & L1$_2$ Sc(Pd,Au)$_{3}$; $Pbam$ Pd/Au order absent \\
17 & \ch{PtNiGe2} & B & -5.8 & & 17.9 & 0.51 & \makecell[tl]{MnP-type (Ni,Pt)Ge, parent of the on-hull structure, \\ Ni/Pt order absent; pattern unresolved} \\
18 & \ch{Gd3Sn6Co}$^{*}$ & B & -25.1 & & & 0.24 & \makecell[tl]{Known CeNiSi$_{2}$-type GdCo$_x$Sn$_{2}$; Co ordering of \\ the on-hull structure absent} \\
\midrule
\multicolumn{8}{l}{\textbf{Other solid solution}} \\
19 & \ch{ScPd2Au} & A & 6.4 & 2078.0 & -4.7 & 0.56 & \makecell[tl]{Hexagonal Sc(Pd,Au)$_{3}$ majority; fcc-stacked Sc(Pd,Au)$_{3}$ minority \\ is the L1$_2$ parent of the on-hull structure} \\
\midrule
\multicolumn{8}{l}{\textbf{Multiphase, on-hull structure not formed}} \\
20 & \ch{ScPd2Cu} & A & 17.7 & 33.2 & 4.8 & 0.75 & \makecell[tl]{fcc-derived Sc(Cu,Pd)$_{3}$-type plus hexagonal \\ phase; compositions undetermined} \\
21 & \ch{La5Al3Co2} & B & -0.7 & & & 0.75 & \makecell[tl]{Known La$_{4}$Co$_{3}$Al$_{3}$ majority; on-hull structure not \\ identified, minority up to $\sim$10 wt\% not excluded} \\
22 & \ch{TiPt5Ge} & B & -18.8 & & -34.0 & 0.20 & \makecell[tl]{Pt + Fe$_{2}$P-type Pt$_{2}$Ge, no Ti-bearing phase \\ identified; on-hull structure absent} \\
23 & \ch{Ti2Pt8Si} & B & -4.8 & & 90.4 & 0.19 & TiPt$_{5}$Si-type and Pt-rich fcc phase formed; on-hull structure absent \\
24 & \ch{LaNiSi3} & C & -24.7 & & -5.2 & 0.07 & Multiphase, known compounds; on-hull structure absent \\
25 & \ch{GdSn6Co4} & B & -4.1 & & -11.2 & 0.04 & \makecell[tl]{Multiphase, known compounds, GdCo$_{6}$Sn$_{6}$ \\ the largest; on-hull structure absent} \\
26 & \ch{NbNi2B2} & C & 12.1 & & 23.0 & 0.03 & \makecell[tl]{Multiphase: Ni$_{3}$B, NbB$_{2}$, $\tau$-boride Nb$_{2}$Ni$_{21}$B$_{6}$; \\ on-hull structure absent} \\
27 & \ch{GdFeGe2} & B & 0.2 & & -6.3 & 0.01 & Two known phases and two minor ones; on-hull structure absent \\
28 & \ch{Gd2Fe3Ge5} & B & -8.6 & & & 0.01 & Multiphase, four known phases; on-hull structure absent \\
29 & \ch{GdFeGe3} & B & -19.0 & & -30.6 & 0.01 & Multiphase, three known phases; on-hull structure absent \\
30 & \ch{LaRuC2} & C & 0.9 & & 28.3 & 0.00 & \makecell[tl]{Ru and probably La(OH)$_3$; \\ on-hull structure absent} \\
\midrule
\multicolumn{8}{l}{\textbf{On-hull structure did not form but structure(s) unresolved}} \\
31 & \ch{Pd6Cu5Ge} & A & -19.6 & & -15.9 & 0.93 & \makecell[tl]{Not $Cmmm$; B2-type when measured on a cut face, \\ broad fcc after grinding; unresolved} \\
32 & \ch{La2PdAu2} & A & -0.8 & & 11.9 & 0.88 & Not the on-hull structure; multiphase, no phase identified \\
33 & \ch{Ti2Pt8Ge} & B & -3.9 & -45.4 & 103.0 & 0.26 & \makecell[tl]{Pt-rich fcc-type phase plus unidentified \\ lines; on-hull structure absent} \\
34 & \ch{ScPdCu4} & A & -24.2 & & 3.9 & 0.25 & Broad pattern, no well-defined lines; on-hull structure absent \\
\bottomrule
\end{tabular}
\end{table*}

\subsection{New compounds and structures}
\ch{GdPt2Ge}, \ch{TiPt5Si} and \ch{ZrPt5Si} formed on-hull structures, and we found no earlier report of any of these compounds in the literature.
Each of these compounds is more than 90~wt\% of its sample after one week at \SI{1000}{\celsius} (Methods).

\ch{GdPt2Ge} formed the YPd$_2$Si-type $Pnma$ structure, which is the lowest-energy structure at this composition in Alexandria.
The $Pnma$ structure is not the lowest-energy one in OQMD, the database from which we selected the target, where a $Cmcm$ structure is lowest.
After annealing, \ch{GdPt2Ge} is 94~wt\% of the sample, and the remaining 6~wt\% is likely \ch{Gd2O3} (Fig.~\ref{fig:xrd_representative}a).

\ch{TiPt5Si} and \ch{ZrPt5Si} formed the predicted $P4/mmm$ structure, which is the lowest-energy structure in every database that lists these compositions.
After annealing, \ch{TiPt5Si} is 91~wt\% of its sample, with the rest an fcc phase, and \ch{ZrPt5Si} is 97~wt\%, with 3~wt\% \ch{ZrPt3} (Fig.~\ref{fig:xrd_representative}b,c).
All three compounds were already identifiable in the as-cast buttons.
As cast, \ch{GdPt2Ge} was already the main phase, and annealing raised the fractions of \ch{TiPt5Si} and \ch{ZrPt5Si}.
Supplementary Fig.~\ref{fig:S_ascast_annealed} compares the as-cast and annealed fits.

The composition \ch{YCoSi3} was listed in a 1974 study of the Y--Co--Si system, which gave the structures of the other ternary compounds of that system but not of \ch{YCoSi3}~\cite{Bodak1974}.
Our sample formed a $Cmmm$ structure of the SmNiGe$_3$ type, \SI{8}{meV/atom} above the predicted $I4mm$ structure in OQMD, and is therefore an above-hull polymorph.
After annealing, the pattern is fitted by the $Cmmm$ phase alone (Fig.~\ref{fig:xrd_representative}d).
To our knowledge, this is the first structure reported for \ch{YCoSi3}.

\ch{YAl3Cu2} formed the predicted CaCu$_5$-type structure, in which Al fills the 3g site and Cu the 2c site (Fig.~\ref{fig:xrd_representative}e).
Al and Cu mostly occupy their predicted sites, and a model with Al and Cu distributed at random fits clearly worse (Supplementary Section~\ref{sec:S_samples}).
The same structure type is known in the Y--Al--Cu system at a different composition, \ch{YAl2Cu3}~\cite{Krachan2003}.
In a single crystal of \ch{YAl2Cu3}, Kadir et al.~\cite{Kadir2001} found Cu alone on the 2c site and Al and Cu sharing the 3g site, the same site preference as in the prediction.
Lines of an unidentified phase remain in the pattern, so we do not quote weight fractions.
\ch{YAl3Cu2} is a nearly ordered compound in the predicted structure at a composition not reported before, but its structure type was already known in this system at a different composition.

\subsection{Previously reported compounds}
\ch{YSnCu}, \ch{ScCrC2} and \ch{YCoSi} formed compounds that had been reported before~\cite{Sebastian2006,Pottgen1995,Rogl1984} (asterisks in Table~\ref{tab:xrd_outcomes}).
In selecting targets, we screened out reported compounds mainly through the databases themselves (Methods).
\ch{YSnCu}, for example, is in the ICSD, but MP labels its entry ``theoretical''~\cite{Jain2013}, so the database flags for experimentally observed compounds are not complete.
We found the papers reporting these three compounds, and the 1974 paper listing the composition \ch{YCoSi3}~\cite{Bodak1974}, only in a later and more thorough search of the literature.
\ch{YSnCu} and \ch{ScCrC2} formed the on-hull structure, \ch{YSnCu} as 90~wt\% of its sample and \ch{ScCrC2} nearly single phase (about 95~wt\%).
\ch{YCoSi} formed its known $Pnma$ structure, \SI{32}{meV/atom} above the predicted $P4/nmm$ structure, as 75~wt\% of its sample, and is therefore an above-hull polymorph.

\subsection{Disordered parents of the on-hull structure}
Ten targets formed a solid solution derived from the on-hull structure in which two or more elements mix on a site (Table~\ref{tab:xrd_outcomes}).
We assign this category when the pattern matches a disordered parent and the order that distinguishes the on-hull structure is absent or cannot be determined.
We judge this from refinements of ordered and disordered models of each target and from the superlattice lines that only the ordered structure produces.
Figure~\ref{fig:xrd_representative}f shows annealed \ch{Hf2MoIr}, whose main phase is a bcc cell in which Hf partly orders onto the cube corners while Mo and Ir share the other site.
In the pattern simulated from the predicted L2$_1$ structure, the order of Mo and Ir produces a (111) line near \SI{23.6}{\degree}, and this line is absent from the data (Fig.~\ref{fig:xrd_representative}f, inset).
The remaining 12~wt\% of the sample is likely a Ti$_2$Ni-type \ch{Hf2Ir} phase.

\begin{figure*}[!t]
    \centering
    \includegraphics[width=0.90\textwidth]{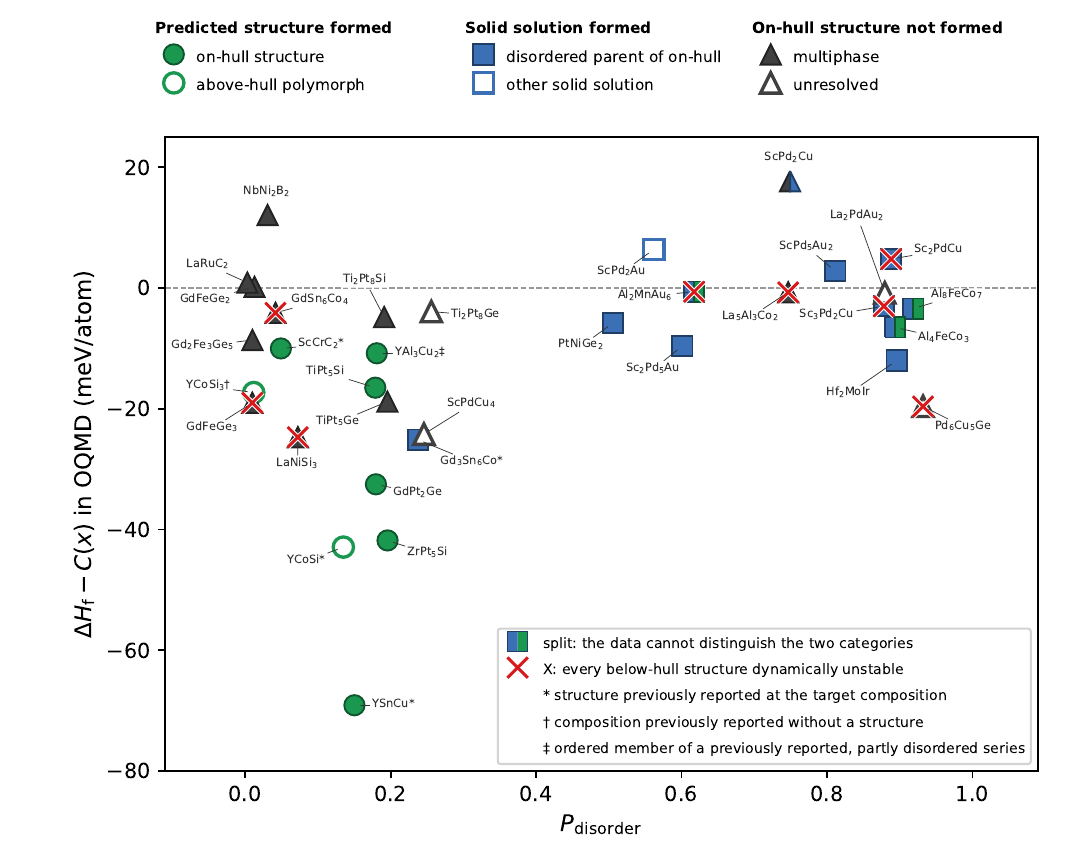}
    \caption{$P_{\text{disorder}}$ versus $\Delta H_\mathrm{f}-C(x)$ in OQMD~\cite{Saal2013,Kirklin2015} for the 34 targets of Table~\ref{tab:xrd_outcomes}. Green circles: the predicted structure formed (filled: on-hull structure, the lowest-energy structure in at least one database that places it below the hull; open: above-hull polymorph). Blue squares: a solid solution formed (filled: disordered parent of the on-hull structure; open: other solid solution). Dark gray triangles: the on-hull structure did not form (filled: multiphase; open: unresolved). A split symbol marks a sample whose data allow two categories that the data cannot distinguish. $^{*}$, $^{\dagger}$ and $^{\ddagger}$ as in Table~\ref{tab:xrd_outcomes}. Red X: every structure below the hull in any of the three databases is dynamically unstable (Methods). Dashed line: $\Delta H_\mathrm{f}-C(x)=0$.}
    \label{fig:disorder_stability}
\end{figure*}

For three targets, our fits cannot distinguish the disordered parent from the on-hull structure.
The patterns of \ch{Al8FeCo7} and \ch{Al4FeCo3} are fitted by a B2 cell with Al on one site and Fe and Co sharing the other, and the predicted structures differ from this disordered model only in how Fe and Co are arranged.
Because Fe and Co differ by one in atomic number, the superlattice lines of Fe/Co order would have at most 0.04\% of the intensity of the strongest line with Cu~K$\alpha$ radiation, and the ordered and disordered models fit equally well (Supplementary Section~\ref{sec:S_samples}).
In \ch{Al2MnAu6}, the predicted $C2/m$ structure and its C11$_b$ parent, in which Al and Mn share a site, also fit the pattern equally well.
We list all three as disordered parents, but each could equally be the on-hull structure, and the split symbols in Fig.~\ref{fig:disorder_stability} mark both possibilities.

\subsection{Other outcomes}
\ch{ScPd2Au} formed a solid solution whose structure is not derived from the on-hull structure (Supplementary Section~\ref{sec:S_samples}).

Eleven targets separated into several phases, none of which is the on-hull structure.
In most of them, the phases we could identify are previously reported compounds.
For example, \ch{GdFeGe3} separated into three known phases (Supplementary Section~\ref{sec:S_samples}).
The pattern of \ch{ScPd2Cu} can also be fitted by a disordered parent of the on-hull structure, so its split symbol in Fig.~\ref{fig:disorder_stability} shows both multiphase and disordered parent.

In \ch{ScPdCu4}, \ch{Pd6Cu5Ge}, \ch{La2PdAu2} and \ch{Ti2Pt8Ge} the on-hull structure does not match the pattern, and we could not identify the phases that formed.
Supplementary Section~\ref{sec:S_samples} gives further details.

\section{Discussion}

Figure~\ref{fig:disorder_stability} plots, for every target, the two quantities that can be evaluated before synthesis: the disorder probability $P_{\text{disorder}}$ and the margin below the hull, $\Delta H_\mathrm{f}-C(x)$ (with the OQMD values).
The symbols mark the outcome categories of Table~\ref{tab:xrd_outcomes}.
A split symbol marks a target whose data allow two categories, such as the three compositions for which ordered and disordered models fit equally well.
Crosses mark targets whose structures below the hull are all dynamically unstable (Methods).
The disorder probability $P_{\text{disorder}}$, from the model of Jakob et al.~\cite{Jakob2025}, is the predicted probability that a compound of the target composition has two or more elements sharing a site.

The targets that formed an on-hull structure or an above-hull polymorph cluster in the lower left of Fig.~\ref{fig:disorder_stability}, where $P_{\text{disorder}}$ is low and the target lies below the hull, and they lie further below the hull than most of the targets that did not form.
None of the 14 targets with $P_{\text{disorder}} > 0.5$ formed, and 9 of them gave a disordered parent of the on-hull structure instead.
These 14 come from Lists~A and~B, which were not pre-screened on $P_{\text{disorder}}$.

Gautier et al.~\cite{Gautier2015} chose their targets by an electron count of 18 per formula unit and a large computed margin below the hull, and most of the compounds they attempted formed in the predicted ordered structure.
Their six single-phase products lie at least \SI{130}{meV/atom} below the hull they computed, about five times the median margin of the eight targets that formed in this study.
Their products also mostly have a low disorder probability in the model of Jakob et al.~\cite{Jakob2025}, below 0.2 for 10 of the 15.
Their selection therefore favored the same two properties that distinguish the targets that formed in this study.

About half of the entries in the ICSD contain some form of site disorder~\cite{Antypov2025,Jakob2025}, whereas the database hulls are built from ordered cells at 0~K.
Configurational entropy favors a disordered phase at high temperature~\cite{Park2026}, and when an arc-melted button is quenched from high temperature, the disorder can be retained.
After a week at \SI{1000}{\celsius}, the predicted order was still absent in \ch{Hf2MoIr} and \ch{Sc3Pd2Cu}.
In these two, the disordered phase may be stable at \SI{1000}{\celsius}, with any ordering transition below that temperature, where diffusion is slow.
Otherwise, the predicted order is either not stable at any temperature or did not form within a week.
Since $P_{\text{disorder}}$ is predicted from the composition alone, it does not say which elements will share a site.
Predicting which elements share which sites in the predicted structure can be computationally intensive~\cite{Angqvist2019,Cerasoli2024}.

A sample that decomposes into several phases implies that the hull was incomplete, that errors in the computed energies were large enough to place an unstable composition on or below it~\cite{Bartel2019}, or that temperature or kinetics, which the 0~K hull leaves out, decided the outcome.
For \ch{Ti2Pt8Ge} and \ch{Ti2Pt8Si}, OQMD and Alexandria list different lowest-energy predicted structures against the same competing phases, and their values of $\Delta H_\mathrm{f}-C(x)$ differ by about \SI{100}{meV/atom}, so at least one of the two values is wrong.
More generally, of the 31 targets listed in more than one database, 14 lie below the hull in one database and above it in another (Table~\ref{tab:xrd_outcomes}).

We also calculated the phonon spectra of the low-energy structures at each target composition with the MatterSim 5M model~\cite{Yang2024MatterSim,Han2026,Togo2023,Togo2023JPCM} (Methods).
In \ch{YCoSi3}, the predicted $I4mm$ structure has imaginary phonon modes, while the $Cmmm$ polymorph that formed, about \SI{8}{meV/atom} higher in OQMD, has none.
In \ch{YCoSi}, however, both the predicted $P4/nmm$ structure and the known $Pnma$ structure that formed are dynamically stable, so for \ch{YCoSi} the phonon calculations do not predict which structure forms.
For eight targets, every structure below the hull is dynamically unstable, and none of these eight formed.
Four of them separated into other phases, including \ch{LaNiSi3} and \ch{GdFeGe3}, despite having a low $P_{\text{disorder}}$ and lying well below the hull.

\subsection{Summary and outlook}
In our study of 34 predicted compounds, taking entries on or near the hulls of OQMD, MP and Alexandria without further criteria rarely gave the predicted compound.
Only 8 of the 34 formed an on-hull structure or an above-hull polymorph, and 3 of these had already been reported experimentally.

Eleven targets have $P_{\text{disorder}} < 0.2$, lie below the OQMD hull and are dynamically stable.
Eight of these formed, and none of the other 23 did.
Applying these criteria requires no new DFT calculations.
Combined with practical requirements, such as suitability for arc melting or another synthesis route, criteria of this kind can raise the predictive value of these databases.

Even by a strict standard, counting only ordered compounds whose composition had not been reported at all, this approach found three new materials, \ch{GdPt2Ge}, \ch{TiPt5Si} and \ch{ZrPt5Si}, and it gave the first reported structure of \ch{YCoSi3}.

The hulls of the large databases are the first stage of a larger effort in materials discovery.
A full prediction would treat each system at finite temperature, as a CALPHAD assessment does~\cite{Lukas2007}, with the liquid, the solid solutions and any order--disorder transitions~\cite{Bartel2018,Park2026}.
Such assessments are not yet practical across the whole of chemical composition space.
Both the database hulls and CALPHAD assessments rest on the candidate structures computed for each system, and both will improve as more of these are computed.
In the meantime, the predictions need experimental tests on many more targets than the 34 in this study.
Since most models, including the disorder model we used, are trained on compounds that have already been made, the strongest tests are on compositions that have never been examined~\cite{Juelsholt2026}.
Reporting the failures of such tests along with the successes~\cite{Akrap2026} would let each attempt improve the hulls and show which criteria accurately predict experimental synthesizability.

\section{Methods}

The database releases we used and the dates on which we retrieved data are recorded in the project repository~\cite{Hamlin_synthesizability}.

\subsection{Candidate selection}
The 34 targets come from three lists, each based on a rough idea, at the time of selection, of what might raise the chance that a predicted compound forms.
We give the criteria to show how the targets were chosen.
List~A contains ternary entries of MP and OQMD within \SI{100}{meV/atom} of the hull, in systems that have reported binary compounds on two of the three binary edges but no reported ternary compound in the Materials Platform for Data Science (MPDS)~\cite{Blokhin2020}, and so might hold undiscovered compounds.
List~B contains entries that OQMD listed on the hull at the time of selection, in systems with binary compounds on all three edges and exactly one reported ternary compound, which we expected to make the hull more reliable.
List~C contains ternary compounds on the PBEsol hull of the Alexandria subset computed by Gao et al.~\cite{Gao2025}, with $P_{\text{disorder}} < 0.2$ to avoid disordered products.
From this list we removed compositions at which OQMD has an entry computed from an ICSD record, compounds containing radioactive or toxic elements (Tc, Pm, Th, Pa, U, Np, Pu, Tl and Ac), and the magnetic rare earths Ce to Yb.
Three List~C targets, \ch{LaRuC2}, \ch{NbNi2B2} and \ch{ScCrC2}, came from a parallel selection with the same filters that also required a predicted superconducting transition temperature above \SI{2}{K} in the same data set.
All targets had to contain elements we had on hand and be suitable for arc melting.

\subsection{Hull energies}
We used OQMD~\cite{Saal2013,Kirklin2015}, MP~\cite{Jain2013} and the PBE release of Alexandria~\cite{Schmidt2024,Cavignac2026}.
For each target and database, we built the convex hull of every entry in the ternary system and its binary and elemental subsystems, excluding entries at the target composition, with pymatgen~\cite{Ong2013}.
We then subtracted the hull energy at the target composition from the formation energy of the lowest-energy entry there.

Supplementary Table~\ref{tab:databases} gives the values for all three databases.
Three List~C targets that lie on the PBEsol hull used for their selection lie 23 to \SI{28}{meV/atom} above the Alexandria PBE hull.

\subsection{Disorder probability}
We evaluated $P_{\text{disorder}}$ with the trained classifier released by Jakob et al.~\cite{Jakob2025}.
The model is a recurrent neural network that takes only the chemical composition as input and returns the probability that a compound of that composition has substitutional disorder.
Because it uses no structural information, it gives one value per composition, regardless of the structure a database predicts.

\subsection{Dynamical stability}
We tested the dynamical stability of every OQMD, MP and Alexandria PBE entry at the 34 target compositions that lies within 50~meV/atom of the hull of the other compositions, 134 entries in 84 distinct structures.
We followed the protocol of the PhononBench benchmark~\cite{Han2026}, with the MatterSim universal interatomic potential (MatterSim-v1.0.0-5M)~\cite{Yang2024MatterSim} and phonopy~\cite{Togo2023,Togo2023JPCM}.
Each structure was relaxed with the FIRE algorithm, with the cell and atomic positions free and the space group fixed, until the residual forces fell below 0.005~eV/\AA.
Force constants were computed from displacements of 0.01~\AA\ in a $2\times2\times2$ supercell of the primitive cell and then symmetrized.
A structure was counted as dynamically unstable if any phonon frequency along a high-symmetry path through the Brillouin zone fell below $-0.1$~THz.
A target is marked by a cross in Fig.~\ref{fig:disorder_stability} if every structure with $\Delta H_\mathrm{f}-C(x) < 0$ in any of the three databases is dynamically unstable.

\subsection{Synthesis}
Samples were melted under Ar in an Edmund B\"uhler MAM-1 compact arc melter.
Air-sensitive elements (Sc, Y, La, Gd and Mn) were cleaned and weighed first, and kept in the transfer chamber of an Ar-filled glovebox while the other elements were weighed, to limit oxidation.
Bulk pieces of the elements were melted together.
Powdered Nb was first pressed into a pellet in a hydraulic press and premelted.
Each button was melted three to five times, for 15--30~s each time, and flipped between melts.
Mass loss was generally below 2\%, and the value for each sample is given in the data repository~\cite{Hamlin_synthesizability}.
Some compositions were made multiple times to correct an error in the weighed stoichiometry or to reduce the mass loss. The repository records each remake and its reason.
Where a composition was measured more than once, as cast, after annealing or as a second sample, Table~\ref{tab:xrd_outcomes} reports the outcome from the pattern whose phases we could identify most completely.

\subsection{Annealing}
Samples to be annealed were wrapped in Ta foil, sealed in silica tubes under Ar, held at \SI{1000}{\celsius} for one week and quenched in water.
Table~\ref{tab:xrd_outcomes} states where an outcome comes from an annealed sample, and the Supplement lists every annealed sample and shows the patterns before and after annealing for the three new compounds.

\subsection{X-ray diffraction and phase identification}
All patterns were measured with Cu~K$\alpha$ radiation, most of them on a Siemens D500 diffractometer, from powders.
Two were measured on bulk pieces on a PANalytical X'Pert MPD, and one on a bulk piece on the D500.
Supplementary Section~\ref{sec:S_xrd} gives the mounts, scan ranges and instrumental line profile.
We first identified the phases by hand in MDI JADE against the PDF-5+ 2026 database~\cite{Kabekkodu2024}, with the predicted structures added.
We then repeated the identification with an automated pipeline of search-match and Rietveld refinement (Supplementary Section~\ref{sec:S_phaseid}).
The automated search compared each pattern with reported structures from the PDF-5+ database and the Crystallography Open Database~\cite{Grazulis2012}, and with every entry of OQMD, MP and Alexandria in the ternary system, the predicted structures included.
We added a fully disordered version of every ordered structure, and random fcc, bcc and hcp cells, in which two or more elements share a site.
Candidate combinations of phases were ranked by how well they matched the pattern, by their energy above the hull and by whether their weight fractions were consistent with the nominal composition.
The most plausible combinations were refined by the Rietveld method~\cite{Rietveld1969} in GSAS-II~\cite{Toby2013}, where the data allowed, with a long-range order parameter $\eta$ that is 1 for the predicted order and 0 for a random distribution.
A target was placed in the on-hull or polymorph category when the phase at the target composition was the largest phase in the refinement.
For each target we compared the refinements of ordered and disordered models and of competing combinations, and the Supplement gives these comparisons for every sample.
The search-match code is deposited, but not its library of PDF-5+ cards.
Every fit of record starts only from structures in OQMD, MP, Alexandria and the Crystallography Open Database, from cells we built, and from six structures entered from their publications.
The GSAS-II project files of the fits of record are deposited with the data~\cite{Hamlin_synthesizability}.

\section{Use of AI tools}
We used Claude (Anthropic) throughout the project, in the versions Opus 4.6, Opus 5.5, Fable 5.1 and Sonnet 5.5 as they became available.
Claude helped write the analysis code in the project repository~\cite{Hamlin_synthesizability}, ran refinements according to our instructions, searched the literature and checked references.
We wrote the main text by hand, and Claude suggested revisions, which we accepted, changed or rejected.
Claude drafted the per-sample descriptions in the Supplement from our analysis of the refinements, and we revised them.
We directed each task, checked the code and its output against the measured data, made all final choices of structural model, fit and interpretation, and take full responsibility for the content.

\section{Data availability}
The data of this study and the analysis code are publicly available: \url{https://doi.org/10.5281/zenodo.19685017}.

\section{Code availability}
The code for the search-match, the Rietveld fits, the figures and the tables of this work is available at \url{https://doi.org/10.5281/zenodo.19685017}.

\bibliography{references}

\section*{Funding Declaration}
This work was supported by the National Science Foundation under award DMR-2522891.

\section*{Author contributions}
G.R.S., B.G., R.G.H., P.J.H.\ and J.J.H.\ conceived the project.
G.D.D., J.G., J.N., J.E., M.C., T.M., J.S.K.\ and G.R.S.\ synthesized the samples, and M.C.\ annealed them.
G.D.D., J.S.K.\ and G.R.S.\ collected the XRD data, and G.D.D., M.J., Z.L.\ and J.J.H.\ analyzed them.
P.P.\ and A.V.\ contributed the disorder-model and dynamic-stability calculations and advised on the use of the databases.
J.J.H.\ developed the analysis code.
G.D.D.\ and J.J.H.\ wrote the manuscript, and all authors reviewed it.

\section*{Competing interests}
The authors declare no competing interests.

\section*{Additional information}
\textbf{Correspondence} and requests for materials should be addressed to James J. Hamlin.

\ifSI
  \clearpage
  \onecolumngrid
  \begin{center}
    {\large\bfseries Supplemental Material}
  \end{center}
  \twocolumngrid
  \setcounter{section}{0}\renewcommand{\thesection}{S\arabic{section}}\renewcommand{\theHsection}{S\arabic{section}}
\renewcommand{\thesubsection}{\thesection.\arabic{subsection}}
\setcounter{figure}{0}\renewcommand{\thefigure}{S\arabic{figure}}\renewcommand{\theHfigure}{S\arabic{figure}}
\setcounter{table}{0}\renewcommand{\thetable}{S\arabic{table}}\renewcommand{\theHtable}{S\arabic{table}}
\setcounter{equation}{0}\renewcommand{\theequation}{S\arabic{equation}}\renewcommand{\theHequation}{S\arabic{equation}}

\providecommand{\new}[1]{{\color{red}#1}}
\providecommand{\old}[1]{\ifdefined\sout{\color{red}\sout{#1}}\else{\color{red}[deleted: #1]}\fi}
\section{Samples}\label{sec:S_buttons}

Table~\ref{tab:S_buttons} lists the 59 arc-melted buttons made for the 34 candidate compositions, with the mass loss of each.
For each candidate, one pattern, the pattern of record, gives the outcome in Table~\ref{tab:xrd_outcomes}, and its fit is the fit of record.
Table~\ref{tab:S_buttons} gives the button of each pattern, whether it was as cast or annealed, the instrument and the mount.
For four candidates, \ch{YAl3Cu2}, \ch{ScCrC2}, \ch{Sc2Pd5Au} and \ch{Pd6Cu5Ge}, this pattern is of a later button of the same composition.
Eight samples were annealed (Methods).
For seven of them, the outcome comes from the annealed sample, and for \ch{Sc2PdCu}, whose annealed pattern was inconclusive, from the as-cast pattern.
Section~\ref{sec:S_samples} shows, for each candidate, the fit of its pattern and the reasoning behind its outcome in Table~\ref{tab:xrd_outcomes}.

\begin{table*}[!htbp]
\centering
\footnotesize
\caption{Arc-melted buttons of the 34 candidates, grouped and numbered as in Table~\ref{tab:xrd_outcomes}. Button: the button whose sample number identifies the candidate in the paper. Earlier attempt: a button of the same composition that this button replaced. Remake: a later button of the same composition. Mass loss in percent in parentheses, with -- where it was not recorded. $^{\dagger}$: not scanned. Pattern of record: the pattern that gives the outcome, as cast or annealed, instrument (D500, Siemens D500; PAN, PANalytical), mount (ZB, powder on a zero-background Si slide; PMMA, powder on a poly(methyl methacrylate) slide; bulk, a piece of the button, measured without grinding). Annealed: \SI{1000}{\celsius} for one week.}
\label{tab:S_buttons}
\vspace{4pt}
\setlength{\tabcolsep}{5pt}
\begin{tabular}{rllllll}
\toprule
\# & Compound & Button & Earlier attempt & Remake & \multicolumn{2}{l}{Pattern of record} \\
\midrule
\multicolumn{7}{l}{\textit{On-hull structure}} \\
1 & \ch{ZrPt5Si} & 0561 (1.0) & 0482 (0.3) &  & 0561 & \parbox[t]{0.25\textwidth}{\raggedright annealed, D500, ZB\strut} \\
2 & \ch{YAl3Cu2} & 0540 (3.0) & 0506 (14.3) & 0615 (2.1) & 0615 & \parbox[t]{0.25\textwidth}{\raggedright annealed, D500, ZB\strut} \\
3 & \ch{GdPt2Ge} & 0480 (0.3) &  &  & 0480 & \parbox[t]{0.25\textwidth}{\raggedright annealed, D500, ZB\strut} \\
4 & \ch{TiPt5Si} & 0493 (1.9) &  &  & 0493 & \parbox[t]{0.25\textwidth}{\raggedright annealed, D500, ZB\strut} \\
5 & \ch{YSnCu} & 0507 (0.4) &  &  & 0507 & \parbox[t]{0.25\textwidth}{\raggedright as cast, D500, PMMA\strut} \\
6 & \ch{ScCrC2} & 0512 (3.0) &  & 0605 (--) & 0605 & \parbox[t]{0.25\textwidth}{\raggedright as cast, D500, PMMA\strut} \\
\midrule
\multicolumn{7}{l}{\textit{Above-hull polymorph}} \\
7 & \ch{YCoSi} & 0538 ($<$0.1) & 0503 ($<$0.1) &  & 0538 & \parbox[t]{0.25\textwidth}{\raggedright as cast, D500, PMMA\strut} \\
8 & \ch{YCoSi3} & 0539 (0.5) & 0504 (2.9) &  & 0539 & \parbox[t]{0.25\textwidth}{\raggedright annealed, D500, ZB\strut} \\
\midrule
\multicolumn{7}{l}{\textit{Disordered parent of on-hull structure}} \\
9 & \ch{Al8FeCo7} & 0535 (0.6) & 0487 (9.4) &  & 0535 & \parbox[t]{0.25\textwidth}{\raggedright as cast, D500, PMMA\strut} \\
10 & \ch{Hf2MoIr} & 0466 (0.6) &  &  & 0466 & \parbox[t]{0.25\textwidth}{\raggedright annealed, D500, ZB\strut} \\
11 & \ch{Al4FeCo3} & 0545 (0.8) & 0486 (3.9) & 0534 (3.3), 0544 (10.9)$^{\dagger}$ & 0545 & \parbox[t]{0.25\textwidth}{\raggedright as cast, D500, PMMA\strut} \\
12 & \ch{Sc2PdCu} & 0474 (0.1) &  &  & 0474 & \parbox[t]{0.25\textwidth}{\raggedright as cast, PAN, bulk\strut} \\
13 & \ch{Sc3Pd2Cu} & 0473 (0.2) &  &  & 0473 & \parbox[t]{0.25\textwidth}{\raggedright annealed, D500, PMMA\strut} \\
14 & \ch{ScPd5Au2} & 0477 (15.6) &  &  & 0477 & \parbox[t]{0.25\textwidth}{\raggedright as cast, PAN, bulk\strut} \\
15 & \ch{Al2MnAu6} & 0479 (2.0) &  &  & 0479 & \parbox[t]{0.25\textwidth}{\raggedright as cast, D500, PMMA\strut} \\
16 & \ch{Sc2Pd5Au} & 0478 (0.5) &  & 0610 (1.0) & 0610 & \parbox[t]{0.25\textwidth}{\raggedright as cast, D500, PMMA\strut} \\
17 & \ch{PtNiGe2} & 0481 (0.2) &  &  & 0481 & \parbox[t]{0.25\textwidth}{\raggedright as cast, D500, PMMA\strut} \\
18 & \ch{Gd3Sn6Co} & 0556 (2.0) & 0489 (4.5) & 0536 (9.5) & 0556 & \parbox[t]{0.25\textwidth}{\raggedright as cast, D500, PMMA\strut} \\
\midrule
\multicolumn{7}{l}{\textit{Other solid solution}} \\
19 & \ch{ScPd2Au} & 0476 (0.2) &  &  & 0476 & \parbox[t]{0.25\textwidth}{\raggedright as cast, D500, PMMA\strut} \\
\midrule
\multicolumn{7}{l}{\textit{Multiphase, on-hull structure not formed}} \\
20 & \ch{ScPd2Cu} & 0469 (1.2) &  &  & 0469 & \parbox[t]{0.25\textwidth}{\raggedright as cast, D500, PMMA\strut} \\
21 & \ch{La5Al3Co2} & 0483 ($<$0.1) &  &  & 0483 & \parbox[t]{0.25\textwidth}{\raggedright as cast, D500, PMMA\strut} \\
22 & \ch{TiPt5Ge} & 0491 (1.4) &  &  & 0491 & \parbox[t]{0.25\textwidth}{\raggedright as cast, D500, PMMA\strut} \\
23 & \ch{Ti2Pt8Si} & 0494 (0.4) &  &  & 0494 & \parbox[t]{0.25\textwidth}{\raggedright as cast, D500, PMMA\strut} \\
24 & \ch{LaNiSi3} & 0514 (2.2) &  &  & 0514 & \parbox[t]{0.25\textwidth}{\raggedright as cast, D500, PMMA\strut} \\
25 & \ch{GdSn6Co4} & 0537 (0.5) & 0490 (11.1) &  & 0537 & \parbox[t]{0.25\textwidth}{\raggedright as cast, D500, PMMA\strut} \\
26 & \ch{NbNi2B2} & 0510 (0.5) &  &  & 0510 & \parbox[t]{0.25\textwidth}{\raggedright as cast, D500, PMMA\strut} \\
27 & \ch{GdFeGe2} & 0555 (1.5) & 0497 (5.4) &  & 0555 & \parbox[t]{0.25\textwidth}{\raggedright as cast, D500, PMMA\strut} \\
28 & \ch{Gd2Fe3Ge5} & 0554 (4.5) & 0496 (1.8) & 0557 (5.0)$^{\dagger}$ & 0554 & \parbox[t]{0.25\textwidth}{\raggedright as cast, D500, PMMA\strut} \\
29 & \ch{GdFeGe3} & 0553 (2.3) & 0495 (2.3) &  & 0553 & \parbox[t]{0.25\textwidth}{\raggedright as cast, D500, PMMA\strut} \\
30 & \ch{LaRuC2} & 0513 (6.3) &  & 0614 (0.6), 0617 (2.2) & 0614 & \parbox[t]{0.25\textwidth}{\raggedright as cast, D500, PMMA\strut} \\
\midrule
\multicolumn{7}{l}{\textit{On-hull structure did not form but structure(s) unresolved}} \\
31 & \ch{Pd6Cu5Ge} & 0465 (0.3) &  & 0568 (1.6), 0608 (1.1) & 0568 & \parbox[t]{0.25\textwidth}{\raggedright as cast, D500, PMMA\strut} \\
32 & \ch{La2PdAu2} & 0475 (0.9) &  &  & 0475 & \parbox[t]{0.25\textwidth}{\raggedright as cast, D500, PMMA\strut} \\
33 & \ch{Ti2Pt8Ge} & 0492 (0.5) &  &  & 0492 & \parbox[t]{0.25\textwidth}{\raggedright as cast, D500, bulk\strut} \\
34 & \ch{ScPdCu4} & 0472 (5.1) &  & 0558 (13.0)$^{\dagger}$, 0566 (4.7), 0613 (4.5) & 0472 & \parbox[t]{0.25\textwidth}{\raggedright as cast, D500, PMMA\strut} \\
\bottomrule
\end{tabular}
\end{table*}

\section{Phase identification}\label{sec:S_phaseid}

We first analyzed every pattern with MDI JADE, used interactively, by search-match against the PDF-5+ 2026 database of the International Centre for Diffraction Data (ICDD)~\cite{Kabekkodu2024} and by whole-pattern fitting.
We also searched the predicted structures of OQMD, MP and Alexandria through the JADE user database, but not in the same search as the PDF cards, and without disordered versions of these structures.
This gave a first outcome for each sample and in some cases identified the main phases, but several patterns were left only partly explained.

Following this manual analysis, we built an automated pipeline that matches each pattern against all of these structures together.
For each ternary system, the library holds the elemental, binary and ternary entries of OQMD, MP and Alexandria, the structures of the system in the Crystallography Open Database (COD)~\cite{Grazulis2012}, and the PDF-5+ cards used in the JADE analysis.
Because a disordered phase can give the same main lines as an ordered predicted structure, the library also holds a fully disordered version of each ordered structure, in which every site carries the average composition.
For many systems it also holds partially disordered versions, in which only some sites mix, with cells relaxed with MatterSim, and solid solutions between two known binaries of the same structure type that share an element, such as (Pt,Ni)Ge or Sc(Pd,Cu), with lattice parameters interpolated by Vegard's law.
It also holds random fcc, bcc and hcp cells at the nominal composition, whose volume per atom is the composition-weighted average of the volumes per atom of the elements.
The calculated pattern of each phase includes the instrumental line profile.
Because computed lattice parameters differ slightly from measured ones, the search recalculates the pattern of each phase with its lattice parameters scaled by up to 2\% and keeps the scaling that best matches the measured pattern.
The candidates are single phases, pairs and triples whose compositions combine to give the nominal composition to within 0.15 in barycentric composition coordinates, and pairs of polymorphs at one composition.
Each combination is fitted to the pattern by non-negative least squares and scored by its residual, its energy above the hull and small penalties for extra phases, trace phases and an off-nominal composition, with weights set in the deposited code.
The 15 best-ranked combinations, chosen to differ in their phases, were refined in GSAS-II~\cite{Toby2013} and ranked by $R_\mathrm{wp}$, and combinations whose refined parameters became unphysical were discarded.
The repository~\cite{Hamlin_synthesizability} contains the search-match code but not the PDF-5+ cards, which may not be redistributed, and the final Rietveld fits start only from open structures (Sec.~\ref{sec:S_xrd}).

The automated search did not reliably find phases below about 10~wt\%.
We added such phases by hand, either from the JADE analysis or by testing likely extra phases, such as oxides, against lines the fit left unexplained, and we kept a phase only where lines of its own, not shared with another phase, are present in the data.
With that check, the final fits determine smaller fractions, such as the 6~wt\% of \ch{Gd2O3} in \ch{GdPt2Ge}.

\section{X-ray diffraction and Rietveld refinement}\label{sec:S_xrd}

All patterns were measured with Cu~K$\alpha$ radiation, all but two on the Siemens D500.
Powder ground or filed from the button was mounted on a PMMA slide or, for most annealed samples, on a zero-background Si slide.
On PMMA slides the powder was fixed with silicone grease or petroleum jelly, and the slide and the grease add a broad hump at low angle, so we started each PMMA-slide refinement above it, mostly near \SI{18}{\degree}.
On zero-background slides a few drops of grease diluted to a few percent in hexane left a thin film, onto which the powder was sprinkled, so these patterns have almost no low-angle hump.
The D500 scans run from 5 to 120 or \SI{150}{\degree} in steps of \SI{0.02}{\degree}, and the two PANalytical scans from about 20 to 120 and 30 to \SI{130}{\degree}.
Three patterns are of solid pieces of a button rather than powders, two of them measured on a PANalytical X'Pert MPD, and their entries in Sec.~\ref{sec:S_samples} describe each.

For the D500, we refined the instrumental line profile and zero offset in GSAS-II~\cite{Toby2013} from a pattern of a quartz standard and held them in every final fit.
Only the as-cast fits of Sec.~\ref{sec:S_ascast} refine a zero offset, in place of the sample displacement.
The PANalytical patterns were fitted with that instrument's profile from a \ch{LaB6} standard.

With the zero offset held, we refined the sample displacement, the height of the powder surface relative to the axis of the diffractometer.
For seven powders on PMMA slides, noted in their entries, it refined to more than about \SI{300}{\micro\metre}.
A thin layer of powder on a flat slide cannot be that far out of position, so we held it at zero for these seven.
Any real displacement then shifts their lines without being modelled, so the lattice parameters of these seven samples are uncertain by up to about 0.5\%.
For thin layers of powder on zero-background slides, noted in their entries, we refined a thin-layer intensity correction, which agrees with the mass of powder on the slide.

A script ran each refinement by adding one group of parameters at a time.
The scale factors with the background, the phase fractions, the main-phase lattice parameters with the sample displacement, the preferred-orientation term and the order parameter were kept whenever their values were physical.
Every other group was kept only if it lowered $R_\mathrm{wp}$ (defined below) by more than 0.05 percentage points and left every value physical, which means positive microstrain and crystallite size, occupancies between 0 and 1 and a March--Dollase ratio between 0.3 and 3.
In order, it refined the scale factors and a Chebyshev background, the phase fractions, and the lattice parameters of the main phase with the sample displacement.
It then refined the lattice parameters of each other phase, the microstrain and crystallite size of the main phase, and the microstrain of the larger other phases.
Where the pattern required them, it added an anisotropic microstrain~\cite{Stephens1999}, a March--Dollase preferred-orientation term~\cite{Dollase1986}, and site occupancies or the single order parameter described below.
It ended with an isotropic atomic displacement parameter for the main phase.

Each lattice parameter could move at most 2\% from its starting value for a structure taken from a database or the literature, and at most 5\% for a disordered parent or a solid solution, whose starting cell is only an estimate.
The script undid any step that moved a lattice parameter beyond this limit.
A predicted structure whose lines cannot be matched within 2\% of its computed lattice parameters therefore counts as absent.

$R_\mathrm{wp}$ is the usual weighted profile residual~\cite{Toby2006}.
Because $R_\mathrm{wp}$ depends on the range, the counting statistics and the background, we compare it only between fits of the same scan over the same range with the same options.

The final fits start only from open structures: entries of OQMD, MP, Alexandria and the COD, cells we built for disordered and solid-solution models, and six structures we entered by hand from the crystallographic tables of their publications, so that the repository holds no copyrighted PDF cards.
Two are CeNiSi$_2$-type phases, \ch{GdFe_{0.37}Ge2} of Jemmali et al.~\cite{Jemmali2010} and GdCo$_x$Sn$_2$ of Gil et al.~\cite{Gil2004}.
The others are \ch{ScCrC2} of P\"ottgen et al.~\cite{Pottgen1995}, \ch{LaNi6Si6} of Pani et al.~\cite{Pani2014}, \ch{La4Co3Al3} of Tursina et al.~\cite{Tursina2023}, and the La$_3$Al$_{11}$-type \ch{Y3(Al,Cu)11} partner of \ch{YAl3Cu2} from the thesis of Krachan~\cite{Krachan2018}.
The disordered parents and the models of partial order were built from the predicted structures.
In a model of partial order, one parameter moves the occupancy of each site linearly from its value in the ordered structure to the average over the sites that mix, that is, from full order to a random distribution.

A candidate counts as formed when the phase at the nominal composition is the largest phase by weight fraction in its fit.
To test for order, we fitted the pattern with the ordered on-hull structure and with its disordered parent, the same structure with only the elements that the order separates mixed on their sites.
At each line that the order adds or changes, we compared the net observed intensity, after subtracting the background, with the calculated intensity, using the counting uncertainty of the observed intensity.
If the ordered model calculates these lines above that uncertainty and they are observed, the order is present.
If they are calculated above the uncertainty but not observed, the ordered structure is absent and the candidate is a disordered parent.
If they are calculated below the uncertainty, as for the Fe/Co order of \ch{Al8FeCo7} and \ch{Al4FeCo3}, the two cannot be told apart, and Fig.~\ref{fig:disorder_stability} marks both.
Section~\ref{sec:S_samples} shows, for each candidate, the fit of its pattern together with one or two fits of alternative models of the same pattern, such as the predicted ordered structure in place of the phase that formed, or a model in which the ordered elements share their sites.
The fitted patterns, refinement logs and GSAS-II project files of every fit in Sec.~\ref{sec:S_samples} are in the repository~\cite{Hamlin_synthesizability}.
\section{As-cast and annealed samples}\label{sec:S_ascast}

Figure~\ref{fig:S_ascast_annealed} compares the fits before and after annealing for the three new compounds.
\ch{GdPt2Ge} was already the main phase as cast, and the $Pnma$ phase alone fits the as-cast pattern once a preferred orientation along [100] is included.
\ch{TiPt5Si} is 62~wt\% of the as-cast button, and its (100) superlattice line near \SI{22.7}{\degree} is clear of the lines of \ch{TiPt8} and has the calculated intensity.
Annealing raised the \ch{TiPt5Si} fraction to 91~wt\%, and that of \ch{ZrPt5Si} from 29 to 97~wt\%.
The as-cast fits have a lower $R_\mathrm{wp}$ than the annealed ones, although their patterns are poorer.
The as-cast powders were on PMMA slides and were refined over a shorter range, and the slide background makes up most of their counts.
Because $R_\mathrm{wp}$ includes the background, it is smaller for such patterns, so it cannot be compared between the two scans of a sample.

\begin{figure*}[!htbp]
\centering
\includegraphics[width=\textwidth]{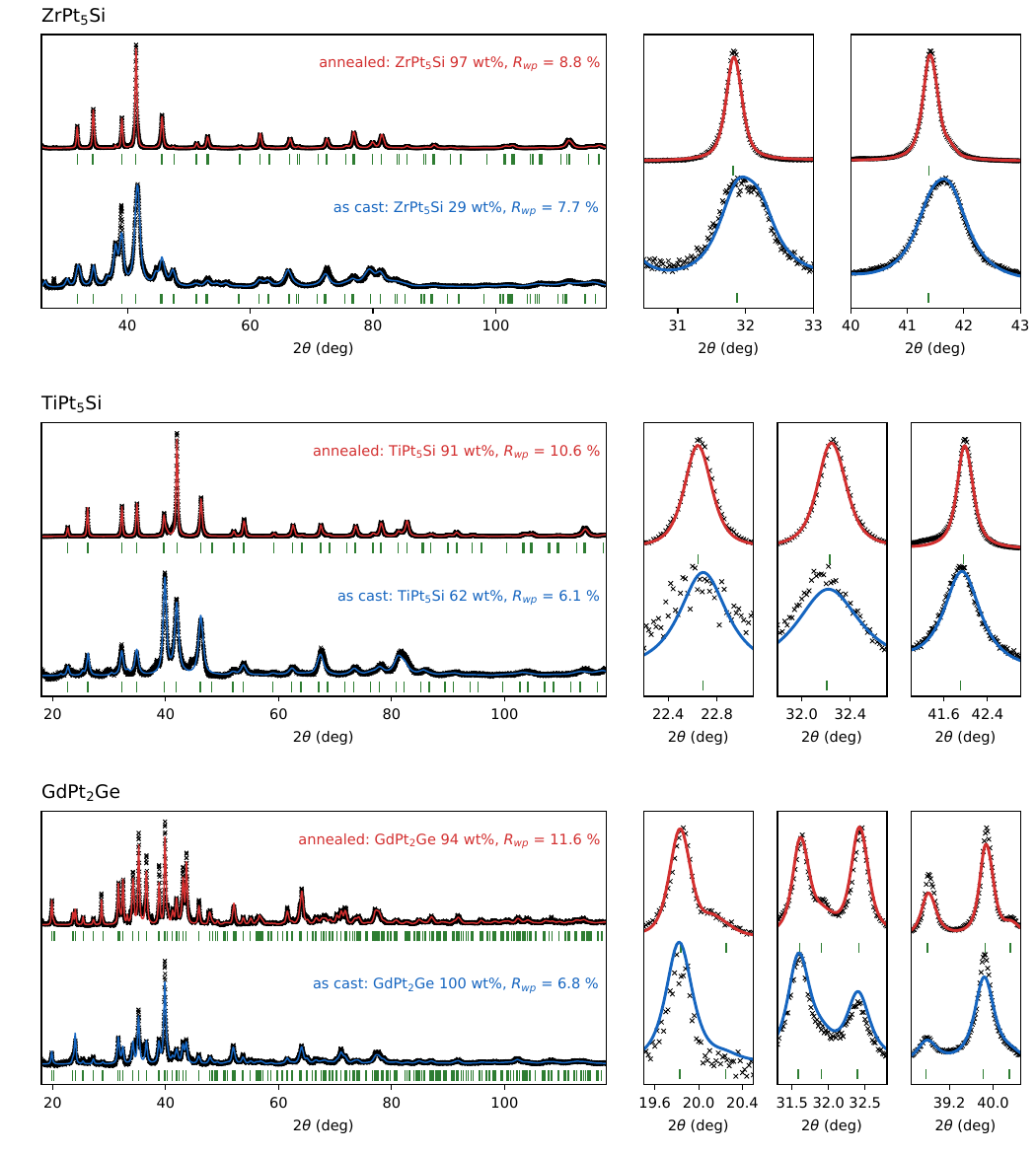}
\caption[Fits of the three new compounds after annealing and as cast]{Fits of the three new compounds after annealing and as cast. Here the fit after annealing is red and the as-cast fit blue (no difference curves are drawn). Left, full patterns, with the measured pattern minus the background (crosses) and the Bragg positions of the phase at the nominal composition (green ticks), labeled with its weight fraction and $R_\mathrm{wp}$. Right, lines of that phase enlarged. $R_\mathrm{wp}$ cannot be compared between the as-cast and annealed scans (Sec.~\ref{sec:S_ascast}). Annealing raised the fractions of \ch{TiPt5Si} and \ch{ZrPt5Si}, while \ch{GdPt2Ge} was already the main phase as cast.}
\label{fig:S_ascast_annealed}
\end{figure*}

\section{Database values}\label{sec:S_databases}

Table~\ref{tab:databases} gives $\Delta H_\mathrm{f}-C(x)$ for each candidate in the three databases, computed as described in Methods.
The values were computed from the entries of each ternary system in OQMD 1.7, the Materials Project (April 2026) and the Alexandria PBE release 2025.07.02.
These entries, extracted from the full databases, are included in the repository~\cite{Hamlin_synthesizability}.

\begin{table*}[!t]
\centering
\footnotesize
\caption{The 34 candidates as seen by the three databases. For each database, $\Delta H_{\mathrm{f}}-C(x)$ is the energy of its lowest entry at the nominal composition relative to the convex hull of all other compositions in that database (meV/atom; negative means the database predicts the compound to be stable), recomputed from the cached entries by the same construction for all three. ``Same'' states whether the database's ground-state structure at that composition is the same structure as the OQMD prediction from which the candidate was selected (pymatgen StructureMatcher); the space group is given where it differs. Blank: no entry at the composition. Grouped and numbered as in Table~\ref{tab:xrd_outcomes}. Snapshots: OQMD v1.7 local dump; Materials Project API, 7 April 2026; Alexandria PBE 2025.07.02 release.}
\label{tab:databases}
\vspace{4pt}
\setlength{\tabcolsep}{5pt}
\begin{tabular}{rllrrlrl}
\toprule
 & & \multicolumn{1}{c}{OQMD} & & \multicolumn{2}{c}{Materials Project} & \multicolumn{2}{c}{Alexandria (PBE)} \\
\cmidrule(lr){3-3}\cmidrule(lr){5-6}\cmidrule(lr){7-8}
\# & Formula & $\Delta H_{\mathrm{f}}-C$ & Structure & $\Delta H_{\mathrm{f}}-C$ & Same & $\Delta H_{\mathrm{f}}-C$ & Same \\
\midrule
\multicolumn{8}{l}{\textit{On-hull structure}} \\
1 & \ch{ZrPt5Si} & -41.8 & $P4/mmm$ & -41.8 & yes & -21.5 & yes \\
2 & \ch{YAl3Cu2} & -10.8 & $P6/mmm$ & & & -3.0 & yes \\
3 & \ch{GdPt2Ge} & -32.5 & $Cmcm$ & & & -25.4 & no ($Pnma$) \\
4 & \ch{TiPt5Si} & -16.5 & $P4/mmm$ & & & -25.6 & yes \\
5 & \ch{YSnCu} & -69.1 & $P6_3mc$ & -69.0 & yes & -69.6 & yes \\
6 & \ch{ScCrC2} & -10.0 & $P6_3/mmc$ & 26.1 & yes & 26.8 & yes \\
\midrule
\multicolumn{8}{l}{\textit{Above-hull polymorph}} \\
7 & \ch{YCoSi} & -42.9 & $P4/nmm$ & 16.9 & no ($Pnma$) & -31.1 & yes \\
8 & \ch{YCoSi3} & -17.3 & $I4mm$ & & & -14.9 & yes \\
\midrule
\multicolumn{8}{l}{\textit{Disordered parent of on-hull structure}} \\
9 & \ch{Al8FeCo7} & -3.4 & $Pm\bar{3}m$ & 12.1 & yes & & \\
10 & \ch{Hf2MoIr} & -12.0 & $Fm\bar{3}m$ & -9.1 & yes & 3.7 & yes \\
11 & \ch{Al4FeCo3} & -6.5 & $Im\bar{3}m$ & & & -3.2 & no ($R\bar{3}m$) \\
12 & \ch{Sc2PdCu} & 4.8 & $Fm\bar{3}m$ & -1.3 & yes & -0.7 & no ($P4/mmm$) \\
13 & \ch{Sc3Pd2Cu} & -3.0 & $P4/mmm$ & & & 4.9 & yes \\
14 & \ch{ScPd5Au2} & 2.8 & $Pbam$ & & & 9.9 & yes \\
15 & \ch{Al2MnAu6} & -0.6 & $C2/m$ & & & 3.2 & yes \\
16 & \ch{Sc2Pd5Au} & -9.6 & $Pbam$ & -12.9 & no ($Amm2$) & -34.7 & yes \\
17 & \ch{PtNiGe2} & -5.8 & $P2_1/m$ & & & 17.9 & no ($P\bar{3}m1$) \\
18 & \ch{Gd3Sn6Co} & -25.1 & $Cmcm$ & & & & \\
\midrule
\multicolumn{8}{l}{\textit{Other solid solution}} \\
19 & \ch{ScPd2Au} & 6.4 & $P4/mmm$ & 2078.0 & no ($Immm$) & -4.7 & yes \\
\midrule
\multicolumn{8}{l}{\textit{Multiphase, on-hull structure not formed}} \\
20 & \ch{ScPd2Cu} & 17.7 & $I4/mmm$ & 33.2 & no ($Fm\bar{3}m$) & 4.8 & no ($P4/mmm$) \\
21 & \ch{La5Al3Co2} & -0.7 & $Cmcm$ & & & & \\
22 & \ch{TiPt5Ge} & -18.8 & $P4/mmm$ & & & -34.0 & yes \\
23 & \ch{Ti2Pt8Si} & -4.8 & $P4/mmm$ & & & 90.4 & no ($Immm$) \\
24 & \ch{LaNiSi3} & -24.7 & $I4mm$ & & & -5.2 & yes \\
25 & \ch{GdSn6Co4} & -4.1 & $I\bar{4}2m$ & & & -11.2 & yes \\
26 & \ch{NbNi2B2} & 12.1 & $I4/mmm$ & & & 23.0 & yes \\
27 & \ch{GdFeGe2} & 0.2 & $P2_1/m$ & & & -6.3 & yes \\
28 & \ch{Gd2Fe3Ge5} & -8.6 & $C2/c$ & & & & \\
29 & \ch{GdFeGe3} & -19.0 & $I4mm$ & & & -30.6 & yes \\
30 & \ch{LaRuC2} & 0.9 & $Cmcm$ & & & 28.3 & no ($Amm2$) \\
\midrule
\multicolumn{8}{l}{\textit{On-hull structure did not form but structure(s) unresolved}} \\
31 & \ch{Pd6Cu5Ge} & -19.6 & $Cmmm$ & & & -15.9 & no ($Immm$) \\
32 & \ch{La2PdAu2} & -0.8 & $C2/m$ & & & 11.9 & yes \\
33 & \ch{Ti2Pt8Ge} & -3.9 & $P4/mmm$ & -45.4 & yes & 103.0 & no ($Immm$) \\
34 & \ch{ScPdCu4} & -24.2 & $F\bar{4}3m$ & & & 3.9 & yes \\
\bottomrule
\end{tabular}
\end{table*}

\section{Dynamical stability}\label{sec:S_phonons}

Table~\ref{tab:S_phonons} lists, for each candidate, the structures that lie below the hull of the other compositions in at least one database, and whether each is dynamically stable.
We relaxed each structure with MatterSim 1.2.5 (5M model) and calculated its phonons with phonopy 4.8.0 by finite displacements in a $2\times2\times2$ supercell.
A structure counts as unstable if any frequency on the seekpath band path lies below \SI{-0.1}{THz}.
Entries of different databases count as one structure when they have the same space group and pymatgen's StructureMatcher matches them.

\begin{table*}[!htbp]
\centering
\footnotesize
\caption{Dynamical stability of the structures below the hull at each candidate composition, grouped and numbered as in Table~\ref{tab:xrd_outcomes}. Each structure is given by its space group, with the databases that contain it (O, OQMD; M, MP; A, Alexandria PBE), in order of increasing $\Delta H_\mathrm{f}-C(x)$. A structure is listed if it lies below the hull of the other compositions in at least one of these databases. Stable: no phonon frequency below \SI{-0.1}{THz} on the band path. $\times$: every structure below the hull is dynamically unstable (crosses in Fig.~\ref{fig:disorder_stability}).}
\label{tab:S_phonons}
\vspace{4pt}
\setlength{\tabcolsep}{5pt}
\begin{tabular}{rlllc}
\toprule
\# & Compound & Dynamically stable & Dynamically unstable & \\
\midrule
\multicolumn{5}{l}{\textit{On-hull structure}} \\
1 & \ch{ZrPt5Si} & \parbox[t]{0.5\textwidth}{\raggedright $P4/mmm$ (O,M,A)\strut} & \parbox[t]{0.2\textwidth}{\raggedright --\strut} & \\
2 & \ch{YAl3Cu2} & \parbox[t]{0.5\textwidth}{\raggedright $P6/mmm$ (O,A)\strut} & \parbox[t]{0.2\textwidth}{\raggedright --\strut} & \\
3 & \ch{GdPt2Ge} & \parbox[t]{0.5\textwidth}{\raggedright $Cmcm$ (O,A), $Pnma$ (A)\strut} & \parbox[t]{0.2\textwidth}{\raggedright --\strut} & \\
4 & \ch{TiPt5Si} & \parbox[t]{0.5\textwidth}{\raggedright $P4/mmm$ (O,A)\strut} & \parbox[t]{0.2\textwidth}{\raggedright --\strut} & \\
5 & \ch{YSnCu} & \parbox[t]{0.5\textwidth}{\raggedright $P6_3mc$ (O,M,A), $F\bar{4}3m$ (A), $Pnma$ (O,A), $P4/nmm$ (A), $P\bar{6}m2$ (A), $Pnma$ (A), $Imm2$ (A)\strut} & \parbox[t]{0.2\textwidth}{\raggedright $P6_3/mmc$ (O)\strut} & \\
6 & \ch{ScCrC2} & \parbox[t]{0.5\textwidth}{\raggedright $P6_3/mmc$ (O,M,A), $Pmmn$ (O,M)\strut} & \parbox[t]{0.2\textwidth}{\raggedright --\strut} & \\
\midrule
\multicolumn{5}{l}{\textit{Above-hull polymorph}} \\
7 & \ch{YCoSi} & \parbox[t]{0.5\textwidth}{\raggedright $P4/nmm$ (O,A), $Pnma$ (O,M,A)\strut} & \parbox[t]{0.2\textwidth}{\raggedright --\strut} & \\
8 & \ch{YCoSi3} & \parbox[t]{0.5\textwidth}{\raggedright $Cmmm$ (O), $P2/m$ (A)\strut} & \parbox[t]{0.2\textwidth}{\raggedright $I4mm$ (O,A)\strut} & \\
\midrule
\multicolumn{5}{l}{\textit{Disordered parent of on-hull structure}} \\
9 & \ch{Al8FeCo7} & \parbox[t]{0.5\textwidth}{\raggedright $Pm\bar{3}m$ (O,M)\strut} & \parbox[t]{0.2\textwidth}{\raggedright --\strut} & \\
10 & \ch{Hf2MoIr} & \parbox[t]{0.5\textwidth}{\raggedright $Fm\bar{3}m$ (O,M,A), $P4/mmm$ (O,A)\strut} & \parbox[t]{0.2\textwidth}{\raggedright --\strut} & \\
11 & \ch{Al4FeCo3} & \parbox[t]{0.5\textwidth}{\raggedright $Im\bar{3}m$ (O), $R\bar{3}m$ (A), $P4/mmm$ (A)\strut} & \parbox[t]{0.2\textwidth}{\raggedright --\strut} & \\
12 & \ch{Sc2PdCu} & \parbox[t]{0.5\textwidth}{\raggedright --\strut} & \parbox[t]{0.2\textwidth}{\raggedright $Fm\bar{3}m$ (O,M,A), $P4/mmm$ (A)\strut} & $\times$ \\
13 & \ch{Sc3Pd2Cu} & \parbox[t]{0.5\textwidth}{\raggedright --\strut} & \parbox[t]{0.2\textwidth}{\raggedright $P4/mmm$ (O,A)\strut} & $\times$ \\
14 & \ch{ScPd5Au2} & \parbox[t]{0.5\textwidth}{\raggedright none below the hull\strut} & \parbox[t]{0.2\textwidth}{\raggedright --\strut} & \\
15 & \ch{Al2MnAu6} & \parbox[t]{0.5\textwidth}{\raggedright --\strut} & \parbox[t]{0.2\textwidth}{\raggedright $C2/m$ (O,A)\strut} & $\times$ \\
16 & \ch{Sc2Pd5Au} & \parbox[t]{0.5\textwidth}{\raggedright $Pbam$ (O,A), $I4/mmm$ (A), $Amm2$ (M)\strut} & \parbox[t]{0.2\textwidth}{\raggedright --\strut} & \\
17 & \ch{PtNiGe2} & \parbox[t]{0.5\textwidth}{\raggedright $P2_1/m$ (O)\strut} & \parbox[t]{0.2\textwidth}{\raggedright --\strut} & \\
18 & \ch{Gd3Sn6Co} & \parbox[t]{0.5\textwidth}{\raggedright $Cmcm$ (O)\strut} & \parbox[t]{0.2\textwidth}{\raggedright --\strut} & \\
\midrule
\multicolumn{5}{l}{\textit{Other solid solution}} \\
19 & \ch{ScPd2Au} & \parbox[t]{0.5\textwidth}{\raggedright $P4/mmm$ (O,A)\strut} & \parbox[t]{0.2\textwidth}{\raggedright --\strut} & \\
\midrule
\multicolumn{5}{l}{\textit{Multiphase, on-hull structure not formed}} \\
20 & \ch{ScPd2Cu} & \parbox[t]{0.5\textwidth}{\raggedright none below the hull\strut} & \parbox[t]{0.2\textwidth}{\raggedright --\strut} & \\
21 & \ch{La5Al3Co2} & \parbox[t]{0.5\textwidth}{\raggedright --\strut} & \parbox[t]{0.2\textwidth}{\raggedright $Cmcm$ (O)\strut} & $\times$ \\
22 & \ch{TiPt5Ge} & \parbox[t]{0.5\textwidth}{\raggedright $P4/mmm$ (O,A)\strut} & \parbox[t]{0.2\textwidth}{\raggedright --\strut} & \\
23 & \ch{Ti2Pt8Si} & \parbox[t]{0.5\textwidth}{\raggedright $P4/mmm$ (O)\strut} & \parbox[t]{0.2\textwidth}{\raggedright --\strut} & \\
24 & \ch{LaNiSi3} & \parbox[t]{0.5\textwidth}{\raggedright --\strut} & \parbox[t]{0.2\textwidth}{\raggedright $I4mm$ (O,A)\strut} & $\times$ \\
25 & \ch{GdSn6Co4} & \parbox[t]{0.5\textwidth}{\raggedright --\strut} & \parbox[t]{0.2\textwidth}{\raggedright $I\bar{4}2m$ (O,A)\strut} & $\times$ \\
26 & \ch{NbNi2B2} & \parbox[t]{0.5\textwidth}{\raggedright none below the hull\strut} & \parbox[t]{0.2\textwidth}{\raggedright --\strut} & \\
27 & \ch{GdFeGe2} & \parbox[t]{0.5\textwidth}{\raggedright $P2_1/m$ (O,A)\strut} & \parbox[t]{0.2\textwidth}{\raggedright --\strut} & \\
28 & \ch{Gd2Fe3Ge5} & \parbox[t]{0.5\textwidth}{\raggedright $C2/c$ (O), $Ibam$ (O)\strut} & \parbox[t]{0.2\textwidth}{\raggedright --\strut} & \\
29 & \ch{GdFeGe3} & \parbox[t]{0.5\textwidth}{\raggedright --\strut} & \parbox[t]{0.2\textwidth}{\raggedright $I4mm$ (O,A)\strut} & $\times$ \\
30 & \ch{LaRuC2} & \parbox[t]{0.5\textwidth}{\raggedright none below the hull\strut} & \parbox[t]{0.2\textwidth}{\raggedright --\strut} & \\
\midrule
\multicolumn{5}{l}{\textit{On-hull structure did not form but structure(s) unresolved}} \\
31 & \ch{Pd6Cu5Ge} & \parbox[t]{0.5\textwidth}{\raggedright --\strut} & \parbox[t]{0.2\textwidth}{\raggedright $Cmmm$ (O,A), $Immm$ (A)\strut} & $\times$ \\
32 & \ch{La2PdAu2} & \parbox[t]{0.5\textwidth}{\raggedright $C2/m$ (O,A)\strut} & \parbox[t]{0.2\textwidth}{\raggedright --\strut} & \\
33 & \ch{Ti2Pt8Ge} & \parbox[t]{0.5\textwidth}{\raggedright $P4/mmm$ (O,M)\strut} & \parbox[t]{0.2\textwidth}{\raggedright --\strut} & \\
34 & \ch{ScPdCu4} & \parbox[t]{0.5\textwidth}{\raggedright $F\bar{4}3m$ (O,A)\strut} & \parbox[t]{0.2\textwidth}{\raggedright --\strut} & \\
\bottomrule
\end{tabular}
\end{table*}

\clearpage
\onecolumngrid
\section{Per-sample analysis}\label{sec:S_samples}
For each candidate, in the order of Table~\ref{tab:xrd_outcomes}, this section gives the fit of record, whose phases define the outcome, and one or two fits of alternative models. Each refines the same scan with the same range and options but changes one model, usually by adding the predicted ordered structure, putting it in place of a phase, or mixing its sites.
In panel (a), crosses show the measured pattern minus the fitted background, the red line the fit of record and the dark blue line the difference.
The rows of tick marks below the pattern, numbered as in the legend, give the Bragg positions of each phase, green for the phase at the nominal composition.
Panel (b) compares the difference curves of the fit of record (red) and of the alternative models (blue dashed, then black dotted), each labeled with its $R_\mathrm{wp}$.
Panel (c) enlarges the deciding lines, marked by gray bands in (a), with each model's calculated pattern in its color, or its stick pattern where its refinement diverged.

\clearpage
\subsection{Samples in which the on-hull structure formed}\label{sec:S_cat_onhull}
\subsubsection{\ch{ZrPt5Si}}\label{sec:S_row_1}

After a week at \SI{1000}{\celsius}, \ch{ZrPt5Si} is 97~wt\% of its sample in the predicted $P4/mmm$ structure, with a little of what is likely \ch{ZrPt3} (Fig.~\ref{fig:S_row_1}).
A model with Zr, Si and Pt mixed on all sites fits far worse ($R_\mathrm{wp} = 23.6\%$ against 8.4\%) and calculates the (110) line at \SI{31.8}{\degree} at one eighth of its observed intensity, so Pt is ordered on its own sites.
Allowing Zr and Si to exchange gives an exchange of about 1\% and no better fit, so Zr and Si are ordered as well.
Every model calculates the (001) line at \SI{12.8}{\degree} too strong, which Zr/Si exchange does not explain.
In the as-cast button, \ch{ZrPt5Si} was a minority phase (Sec.~\ref{sec:S_ascast}).
The prediction is borne out: \ch{ZrPt5Si} formed in the predicted ordered structure.

\begin{figure}[!h]
\centering
\includegraphics[width=\textwidth]{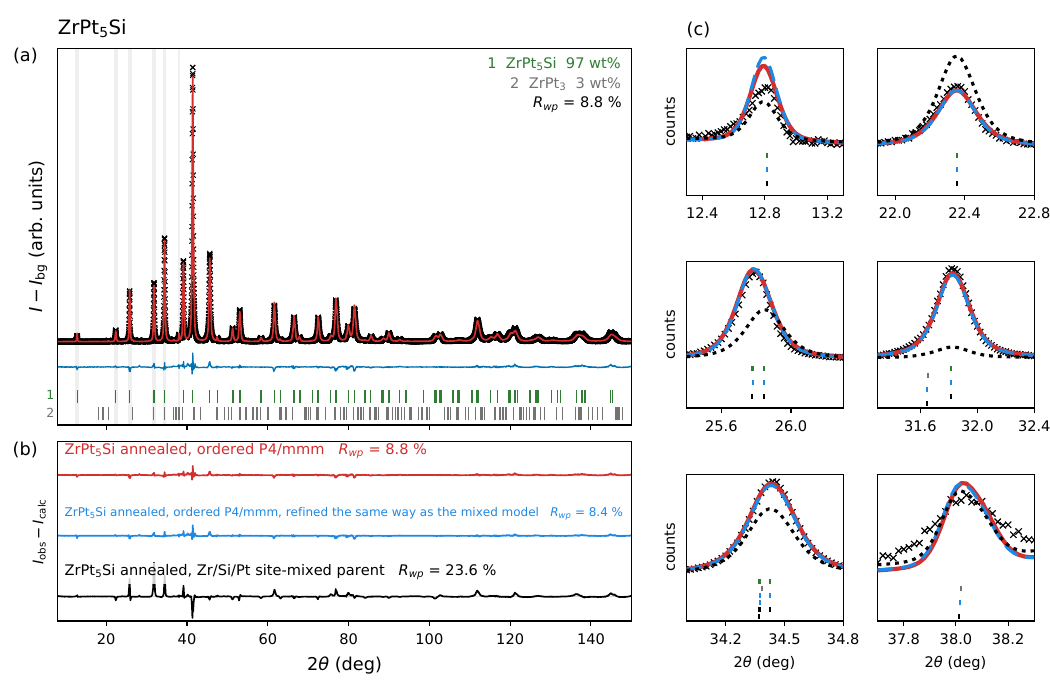}
\caption{Annealed \ch{ZrPt5Si}. In (b) and (c), red is the fit of record, black dotted is the model with Zr, Si and Pt mixed on all sites, and blue dashed is the ordered model refined with exactly the same adjustable parameters as that mixed model, so that blue and black differ only in their site occupancies. At the (110) line near \SI{31.8}{\degree} in (c), the site-mixed model falls far below the data, while the ordered models follow it. At the (001) line near \SI{12.8}{\degree}, all models calculate more intensity than observed.}
\label{fig:S_row_1}
\end{figure}

\clearpage
\subsubsection{\ch{YAl3Cu2}}\label{sec:S_row_2}

The fit of annealed \ch{YAl3Cu2} is poor ($R_\mathrm{wp} = 11.9\%$), with large residuals at the main lines and several lines left unfitted (Fig.~\ref{fig:S_row_2}).
It holds two phases, a CaCu$_5$-type phase in the predicted structure and probably a La$_3$Al$_{11}$-type \ch{Y3(Al,Cu)11} phase~\cite{Krachan2018}, but lines strongest near 29.1, 35.2 and \SI{41.2}{\degree} belong to neither, and some main lines are shifted or have shoulders, so the two phases do not describe the whole sample and we quote no weight fractions.
The conclusion that the predicted Al/Cu order formed rests on one line that the misfit does not affect.
The (001) line of the CaCu$_5$-type phase at \SI{21.7}{\degree} is clearly present in the data.
A model with Al and Cu at random on the 3g and 2c sites gives this line less than a fifth of its observed intensity.
The model with the predicted order, Al on 3g and Cu on 2c, gives the line somewhat more intensity than observed, and a partly ordered model, with Al on 88\% of the 3g sites and Cu on 83\% of the 2c sites, matches the observed intensity.
The predicted structure formed, with Al and Cu mostly on their predicted sites, alongside at least two other phases.

\begin{figure}[!h]
\centering
\includegraphics[width=\textwidth]{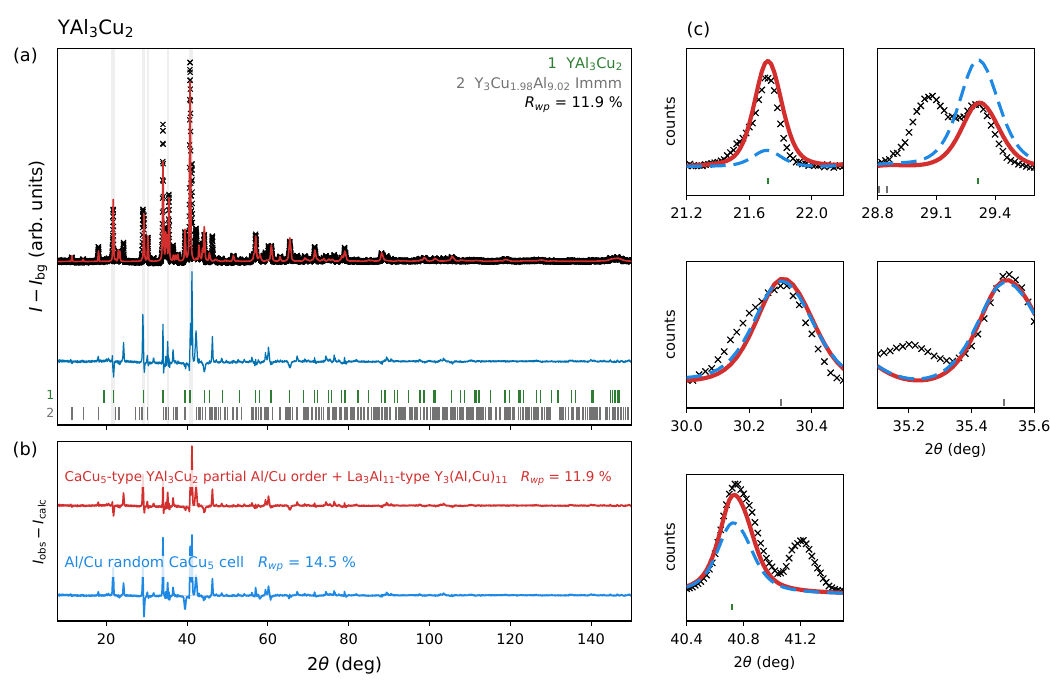}
\caption{Annealed \ch{YAl3Cu2}.
The alternative model in (b) places Al and Cu at random on the 3g and 2c sites of the same CaCu$_5$-type cell.
In (c), only the partly ordered model reproduces the (001) line at \SI{21.7}{\degree}.
The unfitted lines near \SI{29.1}{\degree}, \SI{35.2}{\degree} and \SI{41.2}{\degree} belong to the unidentified phase. Thin-layer intensity correction refined (Sec.~\ref{sec:S_xrd}).}
\label{fig:S_row_2}
\end{figure}

\clearpage
\subsubsection{\ch{GdPt2Ge}}\label{sec:S_row_3}

After annealing, 94~wt\% of the \ch{GdPt2Ge} sample has the YPd$_2$Si-type $Pnma$ structure, and the rest is likely \ch{Gd2O3} (Fig.~\ref{fig:S_row_3}).
This $Pnma$ structure is the lowest-energy structure at this composition in Alexandria.
The $Cmcm$ structure, which OQMD lists as the lowest-energy structure, did not form.

The $Cmcm$ structure does not fit the pattern at all.
Refined alone, it gives $R_\mathrm{wp} = 41\%$, against 14\% for the $Pnma$ structure refined the same way.
It gives no intensity at the strong observed lines at 19.8, 31.6, 32.4, 38.8 and \SI{39.9}{\degree}, and it calculates a line at \SI{26.8}{\degree} where the data show almost nothing.
Two weak lines, at \SI{11.8}{\degree} and \SI{18.0}{\degree}, belong to neither phase.
\ch{GdPt2Ge} formed in the structure that Alexandria predicts as its lowest-energy one, not in the $Cmcm$ structure from OQMD that it was selected from.

\begin{figure}[!h]
\centering
\includegraphics[width=\textwidth]{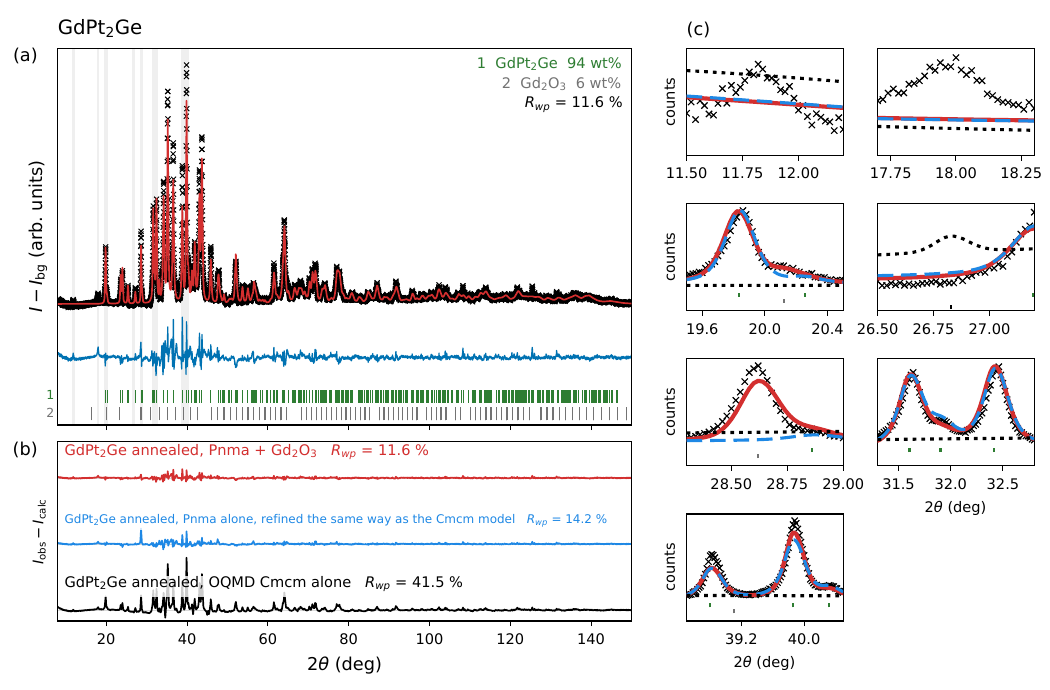}
\caption{Annealed \ch{GdPt2Ge}.
Alternative models in (b): the $Pnma$ phase alone, refined with exactly the same adjustable parameters as the $Cmcm$ model, and the OQMD $Cmcm$ structure alone.
The top two windows in (c) hold the lines at \SI{11.8}{\degree} and \SI{18.0}{\degree} that no model explains.
In the other windows, $Cmcm$ (black dotted) calculates nothing at strong $Pnma$ lines and a line near \SI{26.8}{\degree} that the data lack, and only the fit with the oxide reproduces the \ch{Gd2O3} (222) line.}
\label{fig:S_row_3}
\end{figure}

\clearpage
\subsubsection{\ch{TiPt5Si}}\label{sec:S_row_4}

After annealing, 91~wt\% of the \ch{TiPt5Si} sample has the predicted $P4/mmm$ structure, and the remaining 9~wt\% is an fcc phase (Fig.~\ref{fig:S_row_4}).
A model with Ti, Si and Pt mixed on all sites gives the (110) line at \SI{32.2}{\degree} only a tenth of its observed intensity, so Pt is ordered on its own sites.
This test cannot tell whether Ti and Si are also ordered.
In the as-cast button, \ch{TiPt5Si} was already the main phase (Sec.~\ref{sec:S_ascast}).
The prediction is borne out: \ch{TiPt5Si} formed in the predicted structure, with Pt ordered as predicted.

\begin{figure}[!h]
\centering
\includegraphics[width=\textwidth]{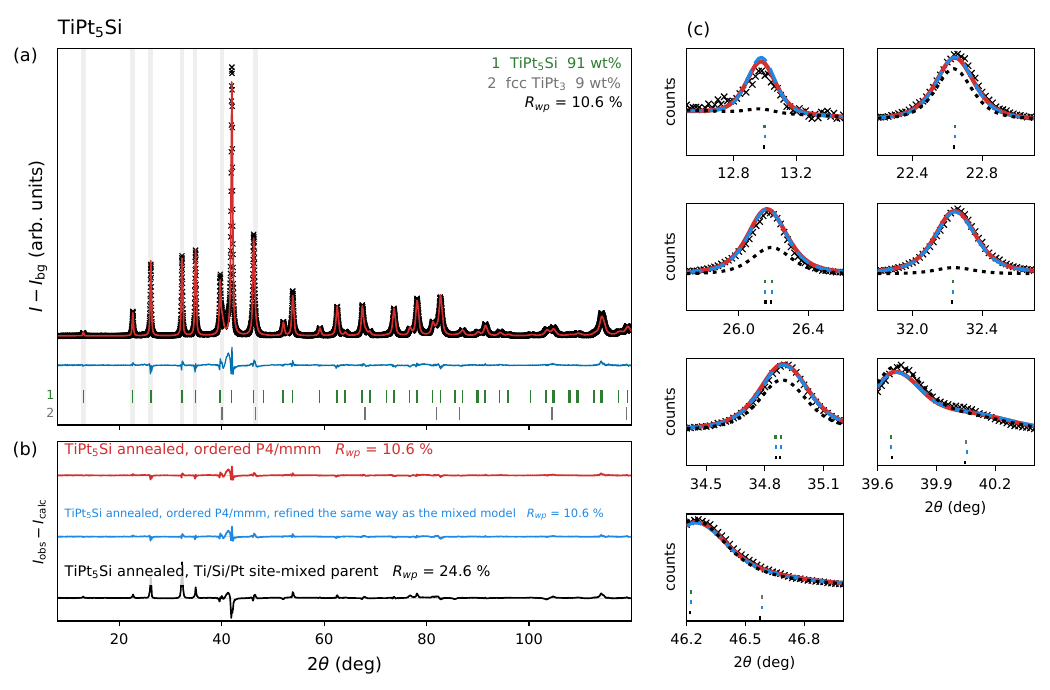}
\caption{Annealed \ch{TiPt5Si}, panels as in Fig.~\ref{fig:S_row_1}.
The alternative model mixes Ti, Si and Pt on every site of the $P4/mmm$ cell.
The alternative model loses the (001) and (110) superstructure lines in (c), and the windows near 40 and \SI{46.5}{\degree} show lines of the fcc phase (gray ticks) next to lines of \ch{TiPt5Si} (green ticks).}
\label{fig:S_row_4}
\end{figure}

\clearpage
\subsubsection{\ch{YSnCu}}\label{sec:S_row_5}

Of the \ch{YSnCu} sample, 90~wt\% has the LiGaGe-type structure reported by Sebastian et al.~\cite{Sebastian2006}, and the rest is likely \ch{Y3Cu4Sn4} (Fig.~\ref{fig:S_row_5}).
This structure is the lowest-energy structure in all three databases.

In this structure Cu and Sn occupy separate sites.
A model with Cu and Sn mixed on both sites fits clearly worse ($R_\mathrm{wp} = 8.7\%$ against 4.9\%) and gives the (101) line at \SI{25.9}{\degree} less than half its observed intensity, so Cu and Sn are ordered.
\ch{YSnCu} formed in the predicted structure, but this compound had been reported before.

\begin{figure}[!h]
\centering
\includegraphics[width=\textwidth]{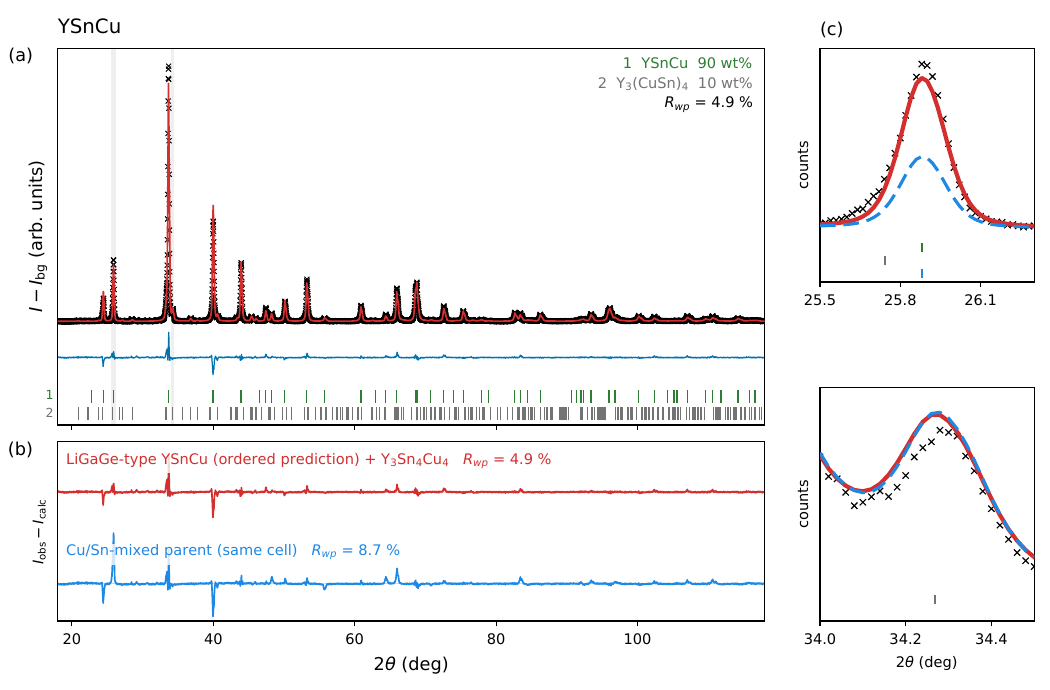}
\caption{As-cast \ch{YSnCu}.
The alternative model in (b) keeps the LiGaGe-type cell and Y site but puts Cu and Sn at random on the two remaining sites.
Its (101) line, in the upper window of (c), is too weak by more than a factor of two.
The lower window shows the strongest isolated line of \ch{Y3Cu4Sn4}.}
\label{fig:S_row_5}
\end{figure}

\clearpage
\subsubsection{\ch{ScCrC2}}\label{sec:S_row_6}

About 95~wt\% of the as-cast \ch{ScCrC2} sample has the $P6_3/mmc$ structure of P\"ottgen et al.~\cite{Pottgen1995}, and the rest is likely graphite (Fig.~\ref{fig:S_row_6}).
This structure is the lowest-energy structure in all three databases.
Sc and Cr differ by only three in atomic number, but their order still shows in the (002) line at \SI{19.4}{\degree}.
A model with Sc and Cr mixed on their sites gives this line less than half its observed intensity.
The first button did not form the carbide, and the pattern shown is from a remake.
A weak line at \SI{35.9}{\degree} is not explained by any phase we tried.
\ch{ScCrC2} formed in the predicted structure, but this compound had been reported before.

\begin{figure}[!h]
\centering
\includegraphics[width=\textwidth]{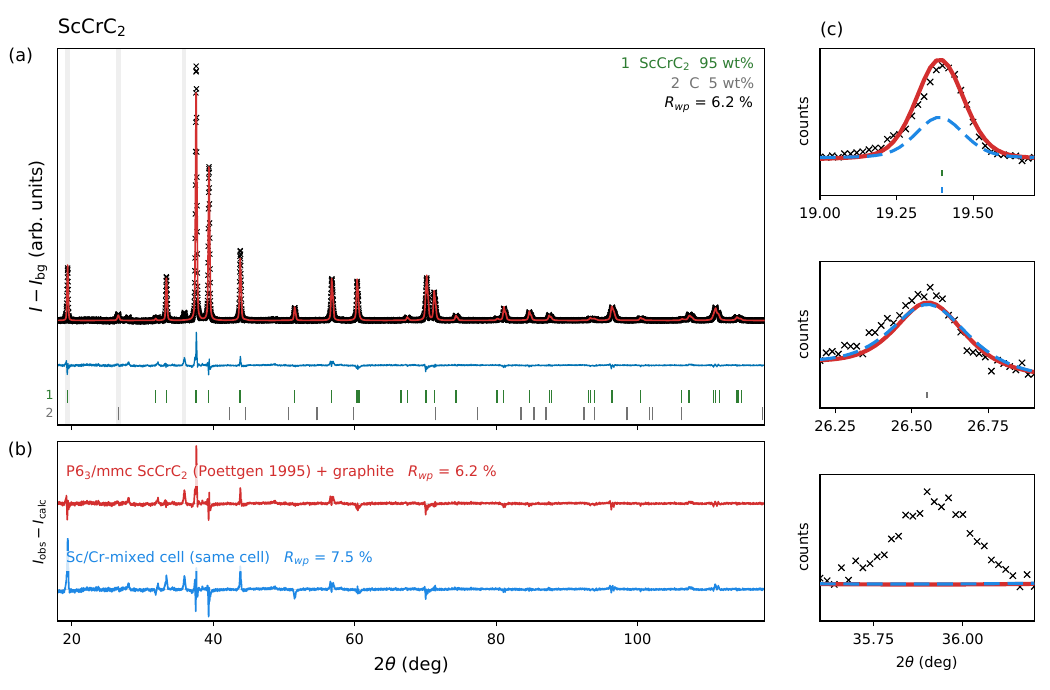}
\caption{As-cast \ch{ScCrC2}.
The alternative model in (b) shares Sc and Cr equally between their two sites.
The windows in (c) show the (002) line of \ch{ScCrC2}, which only the ordered model reproduces, the (002) line of graphite, and the unexplained line at \SI{35.9}{\degree}.}
\label{fig:S_row_6}
\end{figure}

\clearpage
\subsection{Samples that formed an above-hull polymorph}\label{sec:S_cat_abovehull}
\subsubsection{\ch{YCoSi}}\label{sec:S_row_7}

The predicted $P4/nmm$ structure of \ch{YCoSi} did not form.
Instead, 75~wt\% of the as-cast sample has the experimentally known $Pnma$ structure~\cite{Rogl1984}, which OQMD places \SI{32}{meV/atom} above the predicted structure (Fig.~\ref{fig:S_row_7}).
The rest is likely \ch{Y3Co2Si3} and \ch{YCo2Si2}.
The strongest line of the pattern, at \SI{36.2}{\degree}, belongs to the $Pnma$ phase, and the $P4/nmm$ structure gives almost no intensity there.
With the $P4/nmm$ structure in place of the $Pnma$ phase, the fit is far worse ($R_\mathrm{wp} = 40\%$ against 10\%), the refined fraction of the $P4/nmm$ phase drops to about 2~wt\%, and its strong (200) line near \SI{46.1}{\degree} is absent from the data.
The $Pnma$ phase needs a large preferred-orientation correction, so its line intensities are less certain than its line positions.
\ch{YCoSi} formed in a known structure that is higher in energy than the predicted one.

\begin{figure}[!h]
\centering
\includegraphics[width=\textwidth]{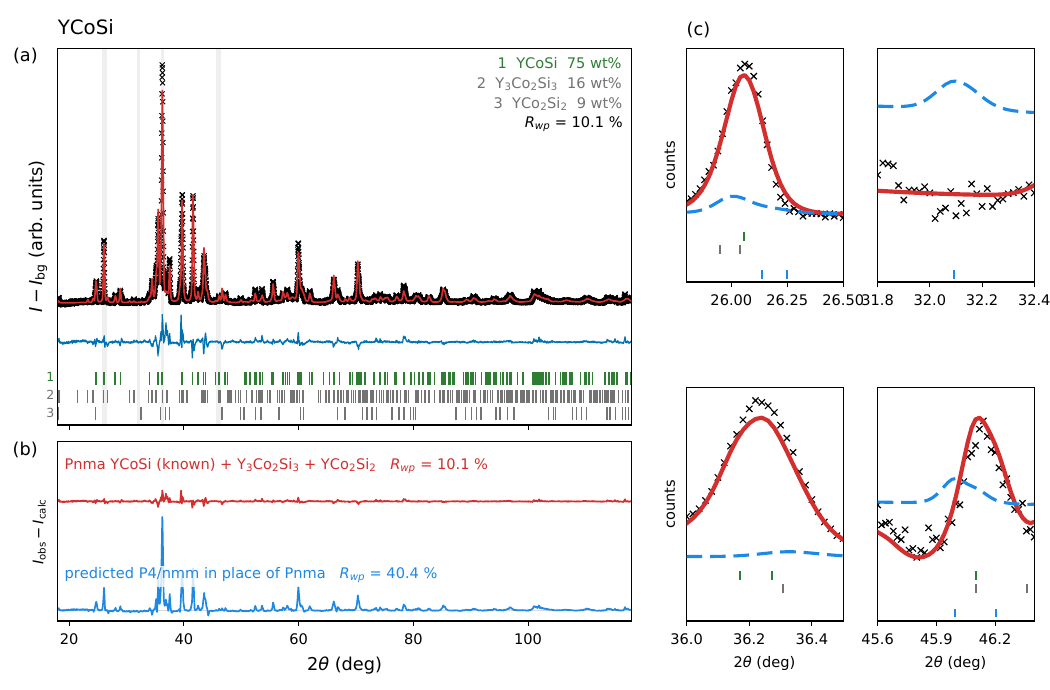}
\caption{As-cast \ch{YCoSi}. In (b) and (c), red is the fit with the experimentally known $Pnma$ phase and blue dashed the fit with the predicted $P4/nmm$ phase in its place, both with the two minor phases. The fit with the predicted phase misses the strongest line, at \SI{36.2}{\degree}, puts intensity near \SI{32.1}{\degree} where the data have none, and adds its (200) line beside the weak $Pnma$ line near \SI{46.1}{\degree}.}
\label{fig:S_row_7}
\end{figure}

\clearpage
\subsubsection{\ch{YCoSi3}}\label{sec:S_row_8}

After annealing, the \ch{YCoSi3} sample is a single phase with the SmNiGe$_3$-type $Cmmm$ structure (Fig.~\ref{fig:S_row_8}).
OQMD places this structure \SI{8}{meV/atom} above the predicted $I4mm$ structure, which it lists as the lowest-energy one.
Bodak et al.~\cite{Bodak1974} reported the composition but no structure.
Two strong lines of the $I4mm$ structure, at \SI{35.9}{\degree} and \SI{36.7}{\degree}, would fall where the data are flat.
A fit of the $I4mm$ structure alone diverged, so the figure shows the positions of its lines instead.
\ch{YCoSi3} formed in a structure that OQMD places slightly above the predicted one, not in the predicted structure.

\begin{figure}[!h]
\centering
\includegraphics[width=\textwidth]{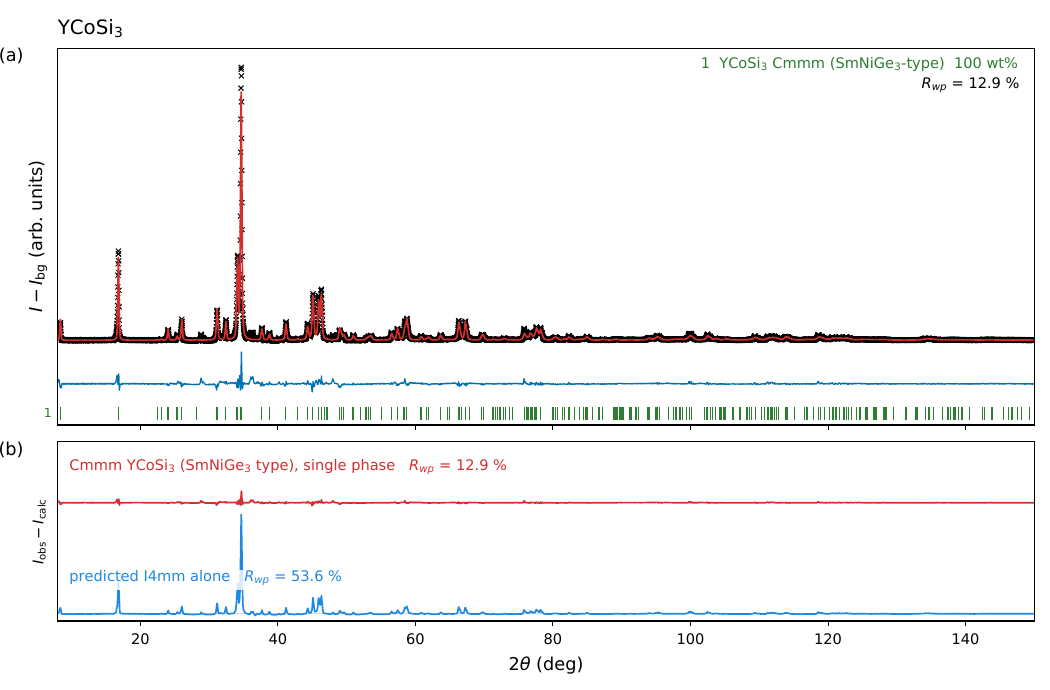}
\caption{Annealed \ch{YCoSi3}. In (b), red is the fit with the $Cmmm$ structure alone and blue the fit with the predicted $I4mm$ structure alone. The $I4mm$ structure fails at every strong line, so its difference curve shows the whole measured pattern. Thin-layer intensity correction refined (Sec.~\ref{sec:S_xrd}).}
\label{fig:S_row_8}
\end{figure}

\clearpage
\subsection{Samples that formed a disordered parent of the on-hull structure}\label{sec:S_cat_disordered}
\subsubsection{\ch{Al8FeCo7}}\label{sec:S_row_9}

\ch{Al8FeCo7} formed a B2 phase, with Al on one site and Fe and Co mixed on the other.
The predicted structure is a $2\times2\times2$ B2 supercell with Fe and Co on separate sites.
X-rays scatter almost equally from Fe and Co, so the lines produced by Fe/Co order are calculated at about 0.01\% of the strongest line, far below the noise.
The ordered model and the model with Fe and Co mixed both give $R_\mathrm{wp} = 5.9\%$, so our data cannot tell whether Fe and Co are ordered.
\ch{Al8FeCo7} formed the predicted lattice with Al ordered, but our data cannot tell whether Fe and Co are also ordered as predicted, so both readings remain possible.

\begin{figure}[!h]
\centering
\includegraphics[width=\textwidth]{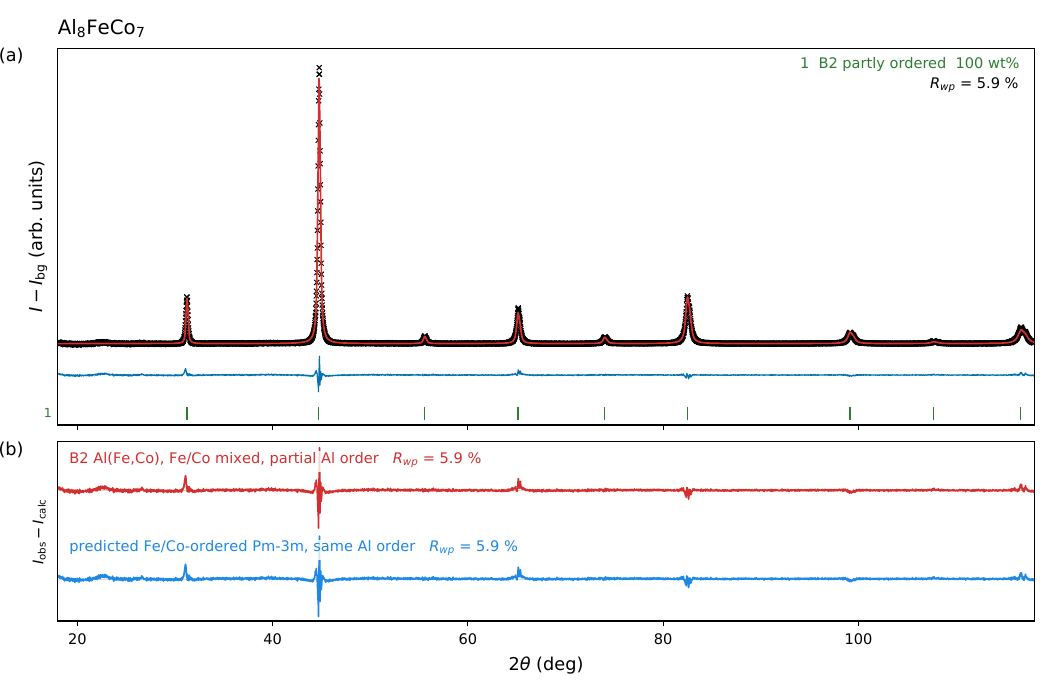}
\caption{As-cast \ch{Al8FeCo7}.
(b) The B2 model with Fe and Co mixed (red) and the predicted Fe/Co-ordered $Pm\bar{3}m$ model with the same Al order (blue) leave the same difference curve.
The Fe/Co superlattice lines are too weak to show, so the figure has no panel (c).}
\label{fig:S_row_9}
\end{figure}

\clearpage
\subsubsection{\ch{Hf2MoIr}}\label{sec:S_row_10}

After annealing, 88~wt\% of the \ch{Hf2MoIr} sample is a bcc phase with partial Hf order, and 12~wt\% is likely Ti$_2$Ni-type \ch{Hf2Ir}.
The predicted L2$_1$ structure also orders Mo and Ir, which produces a (111) line at \SI{23.6}{\degree} calculated at 4\% of the (110) line.
The data show no line there (Fig.~\ref{fig:S_row_10}c).
With the same \ch{Hf2Ir} phase, the L2$_1$ model gives $R_\mathrm{wp} = 16.9\%$, against 7.3\% for the fit of record.
The B2 (100) line at \SI{27.3}{\degree} has less than half the intensity calculated for full Hf order, so the Hf order is partial.
\ch{Hf2MoIr} formed the lattice of the predicted structure, but without the predicted Mo/Ir order and with only partial Hf order.

\begin{figure}[!h]
\centering
\includegraphics[width=\textwidth]{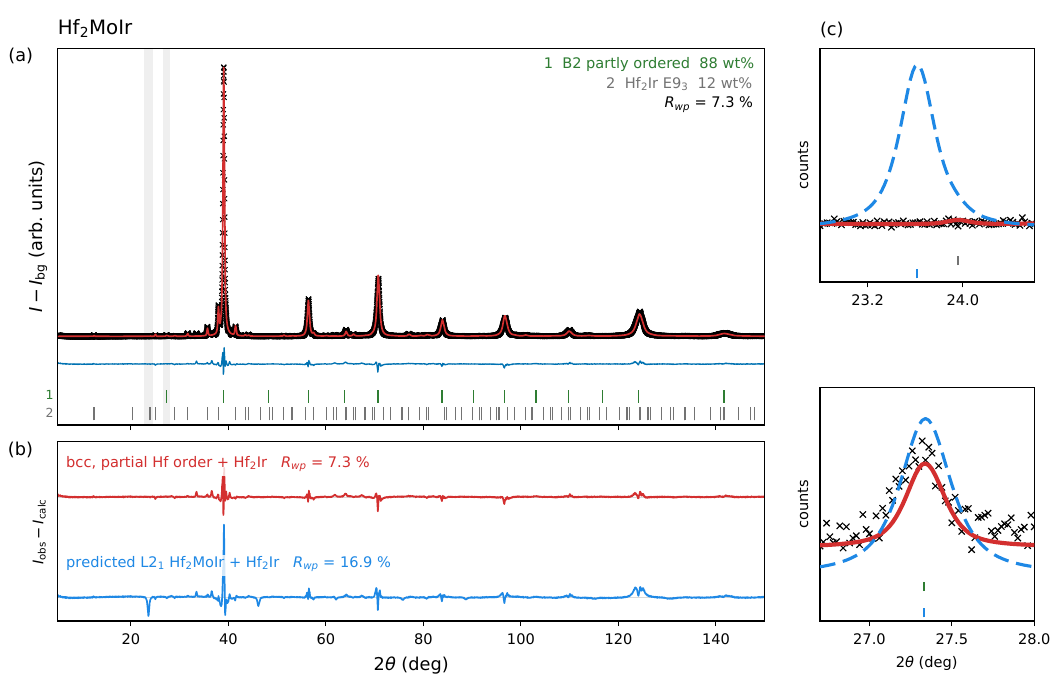}
\caption{Annealed \ch{Hf2MoIr}.
(b) The bcc model with partial Hf order (red) and the predicted L2$_1$ structure (blue), each with \ch{Hf2Ir}.
(c) The L2$_1$ (111) line at \SI{23.6}{\degree}, calculated by the prediction and absent from the data, and the B2 (100) line at \SI{27.3}{\degree}, which full Hf order over-calculates.}
\label{fig:S_row_10}
\end{figure}

\clearpage
\subsubsection{\ch{Al4FeCo3}}\label{sec:S_row_11}

\ch{Al4FeCo3} formed a B2 phase, with Al on one site and Fe and Co mixed on the other.
The predicted structure is a body-centered $Im\bar{3}m$ supercell with Fe and Co on separate sites.
As in \ch{Al8FeCo7}, the lines produced by Fe/Co order are far too weak to detect with X-rays, and the ordered and Fe/Co-mixed models both give $R_\mathrm{wp} = 5.8\%$.
Neutron diffraction could settle the Fe/Co order in both compounds, since the neutron scattering lengths of Fe and Co differ by almost a factor of four.
\ch{Al4FeCo3} formed the predicted lattice with Al ordered, but our data cannot tell whether Fe and Co are also ordered as predicted, so both readings remain possible.

\begin{figure}[!h]
\centering
\includegraphics[width=\textwidth]{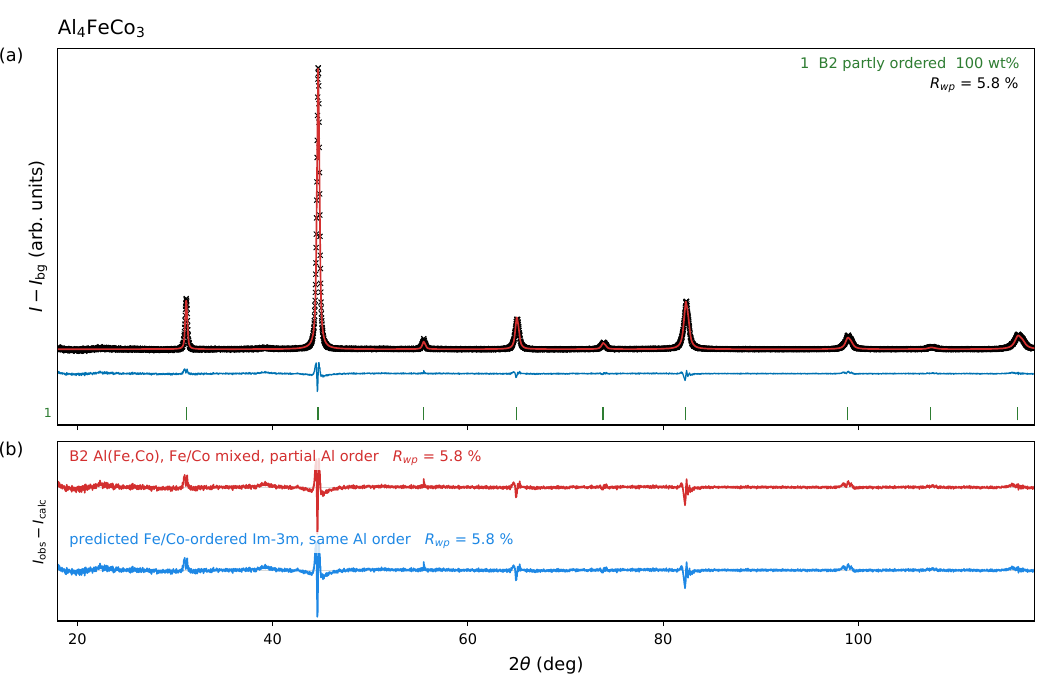}
\caption{As-cast \ch{Al4FeCo3}.
(b) The B2 model with Fe and Co mixed (red) and the predicted Fe/Co-ordered $Im\bar{3}m$ model with the same Al order (blue), which leave the same difference curve.
As for \ch{Al8FeCo7}, the Fe/Co superlattice lines are too weak to show.}
\label{fig:S_row_11}
\end{figure}

\clearpage
\subsubsection{\ch{Sc2PdCu}}\label{sec:S_row_12}

The predicted structures do not match the pattern of \ch{Sc2PdCu}.
The lowest-energy structures, L2$_1$ in OQMD and MP and $P4/mmm$ in Alexandria, both order Cu and Pd, which adds superlattice lines to the B2 pattern.
The L2$_1$ (111) line at \SI{23.6}{\degree} is calculated at 8\% of the strongest line, and the $P4/mmm$ (101) line at \SI{30.7}{\degree} at 4\%, but the data are flat at both positions (Fig.~\ref{fig:S_row_12}c).
The best match we found is a B2 phase, with Sc on one site and Cu and Pd mixed on the other ($R_\mathrm{wp} = 8.1\%$).
The pattern is from a cut face of the as-cast button.
The B2 (100) line at \SI{27.2}{\degree} is weaker than full Sc order calculates, so the Sc order is probably incomplete.
So the predicted Cu/Pd order did not form.

\begin{figure}[!h]
\centering
\includegraphics[width=\textwidth]{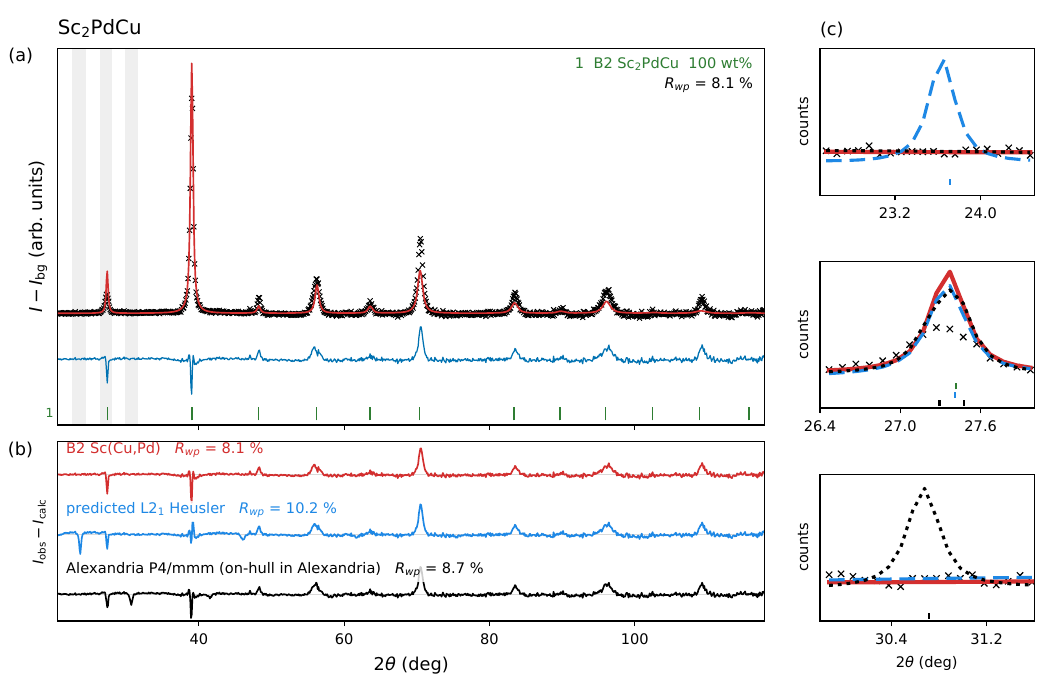}
\caption{As-cast \ch{Sc2PdCu}, bulk cut face.
(b) B2 with full Sc order (red), the predicted L2$_1$ Heusler structure (blue) and the Alexandria $P4/mmm$ structure (black).
(c) The L2$_1$ (111) line at \SI{23.6}{\degree}, the B2 (100) line at \SI{27.2}{\degree} and the $P4/mmm$ (101) line at \SI{30.7}{\degree}.
The ordered structures each calculate a line that the data lack.}
\label{fig:S_row_12}
\end{figure}

\clearpage
\subsubsection{\ch{Sc3Pd2Cu}}\label{sec:S_row_13}

The predicted structure does not match the pattern of \ch{Sc3Pd2Cu}.
It has Cu and Pd in alternate layers, which produces a (001) line at \SI{8.95}{\degree} calculated at 6\% of the strongest line.
We started this refinement at \SI{8}{\degree} to include that line, and the line is absent, even after a week at \SI{1000}{\celsius} (Fig.~\ref{fig:S_row_13}c).
The best match we found is a B2 phase, with Sc partly ordered and Cu and Pd mixed ($R_\mathrm{wp} = 4.3\%$).
So the predicted Cu/Pd order did not form, although the B2 phase has the same underlying lattice as the predicted structure.

\begin{figure}[!h]
\centering
\includegraphics[width=\textwidth]{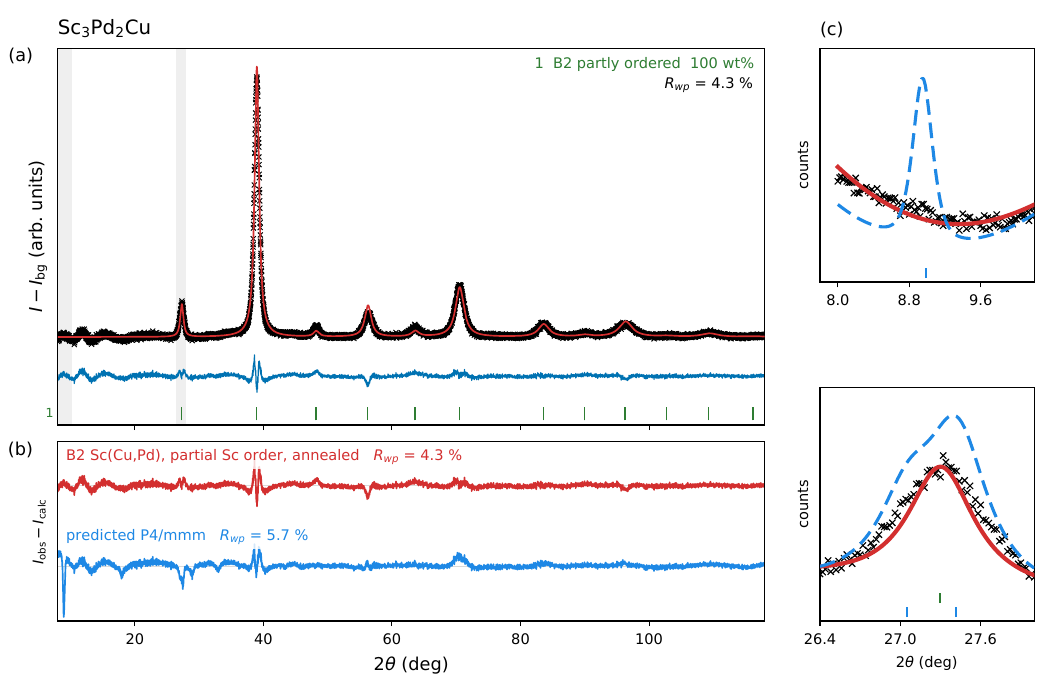}
\caption{Annealed \ch{Sc3Pd2Cu}.
(b) B2 with partial Sc order (red) and the predicted $P4/mmm$ structure (blue).
(c) The $P4/mmm$ (001) line at \SI{8.95}{\degree} and the B2 (100) line at \SI{27.2}{\degree}.
The blue curve includes that fit's own background, which bends below the data around the missing line.}
\label{fig:S_row_13}
\end{figure}

\clearpage
\subsubsection{\ch{ScPd5Au2}}\label{sec:S_row_14}

The fcc lattice matches every strong line of \ch{ScPd5Au2}, but the intensities of several lines are poorly fitted ($R_\mathrm{wp} = 15.3\%$), because the pattern is from the uncut face of the as-cast button rather than a powder (Fig.~\ref{fig:S_row_14}).
The predicted $Pbam$ structure is an ordered version of the same fcc lattice.
Within the measured range it differs from the disordered fcc model mainly by weak superlattice lines, the strongest near \SI{32}{\degree}, calculated at about 5\% of the strongest line.
The data show at most a weak bump near \SI{31.6}{\degree}, slightly below the calculated position, so the conclusion rests on this one weak feature.
The scan starts at \SI{30.5}{\degree}, so the stronger superlattice lines at lower angle were not measured, and the button lost 16\% of its mass during melting.
We take the predicted order as not observed, but it is not excluded, and the conclusion for this sample is provisional.

\begin{figure}[!h]
\centering
\includegraphics[width=\textwidth]{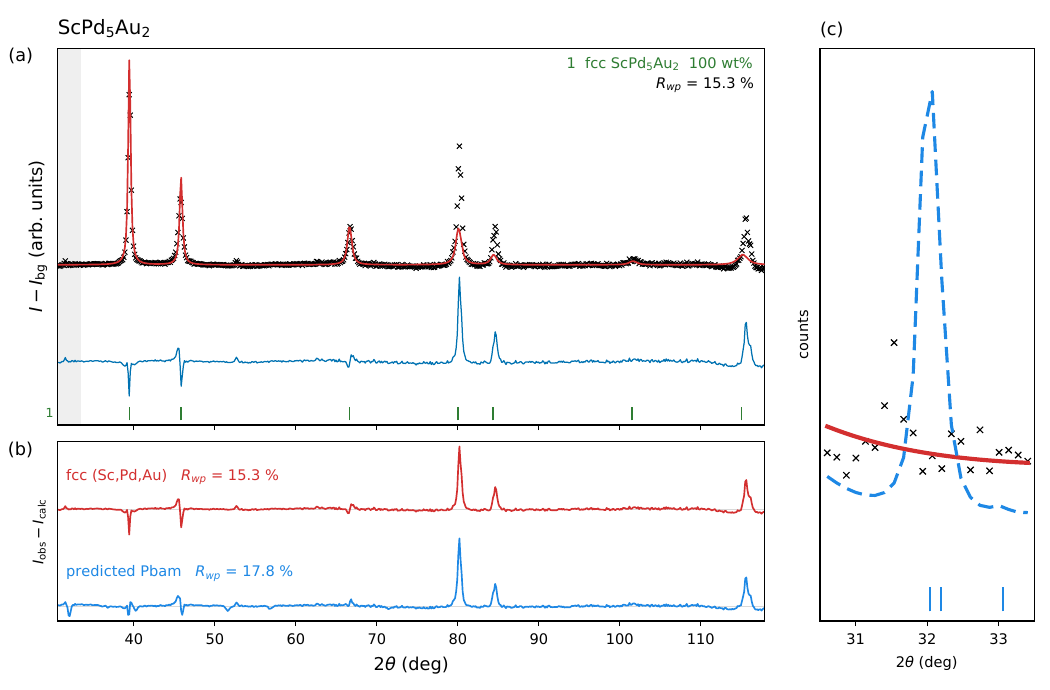}
\caption{As-cast \ch{ScPd5Au2}, uncut button face. In (b) and (c), red is the disordered fcc model and blue dashed the predicted $Pbam$ structure. (c) The region of the strongest $Pbam$ superlattice line in the measured range, near \SI{32}{\degree}, where the data show at most a weak bump near \SI{31.6}{\degree}.}
\label{fig:S_row_14}
\end{figure}

\clearpage
\subsubsection{\ch{Al2MnAu6}}\label{sec:S_row_15}

The best fit to \ch{Al2MnAu6} uses a C11$_b$ cell in which Au is only partly ordered, so that a third of the site shared by Al and Mn also holds Au.
The predicted $C2/m$ structure orders Mn and Al on that site.
With the same partial Au order, the $C2/m$ model fits no worse ($R_\mathrm{wp} = 5.9\%$ against 6.2\%).
The lines produced only by Mn/Al order, near \SI{32.3}{\degree}, are calculated at about 10 counts, well below the scatter of the data.
The small improvement of the $C2/m$ model comes from the extra freedom of its lower, monoclinic symmetry in fitting the broad lines, not from these lines.
Our data therefore cannot tell whether Mn and Al in \ch{Al2MnAu6} are ordered as predicted, so both readings remain possible.

\begin{figure}[!h]
\centering
\includegraphics[width=\textwidth]{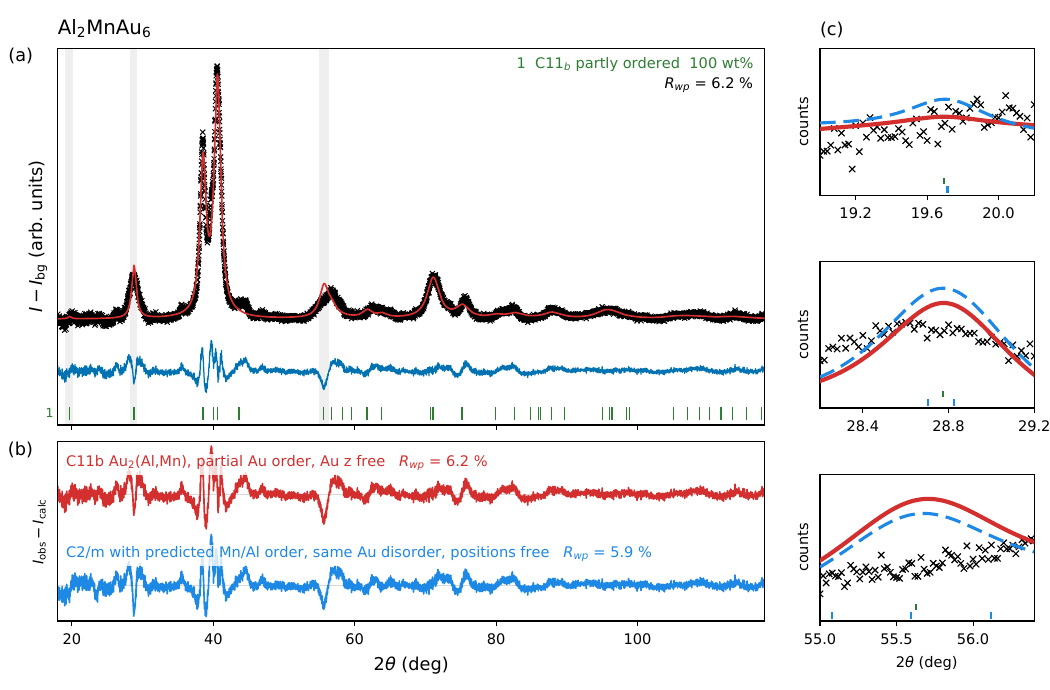}
\caption{As-cast \ch{Al2MnAu6}.
(b) C11$_b$ with Al and Mn mixed (red) and $C2/m$ with the predicted Mn/Al order (blue), both with the same partial Au order.
(c) The C11$_b$ (002) and (101) lines near \SI{19.6}{\degree} and \SI{28.7}{\degree} and the broad line near \SI{55.6}{\degree}.
The two models misplace the lines at \SI{28.7}{\degree} and \SI{55.6}{\degree} by the same amount. Sample displacement held at zero (Sec.~\ref{sec:S_xrd}).}
\label{fig:S_row_15}
\end{figure}

\clearpage
\subsubsection{\ch{Sc2Pd5Au}}\label{sec:S_row_16}

The predicted structures do not match the pattern of \ch{Sc2Pd5Au}.
The $Pbam$ structure, the lowest-energy structure in OQMD and Alexandria, orders Pd and Au, which produces weak lines near \SI{15.7}{\degree} and \SI{17.6}{\degree}.
These lines are calculated at five to ten times the noise, but neither appears in the pattern (Fig.~\ref{fig:S_row_16}c).
The $Amm2$ structure, the lowest-energy structure in MP, does not match either, since its strong lines at \SI{17.9}{\degree} and \SI{26.4}{\degree} fall where the data are flat.
The best match we found is an L1$_2$ Sc(Pd,Au)$_3$ phase, with Sc on the cube corners and Pd and Au mixed on the face centers ($R_\mathrm{wp} = 7.2\%$).

\begin{figure}[!h]
\centering
\includegraphics[width=\textwidth]{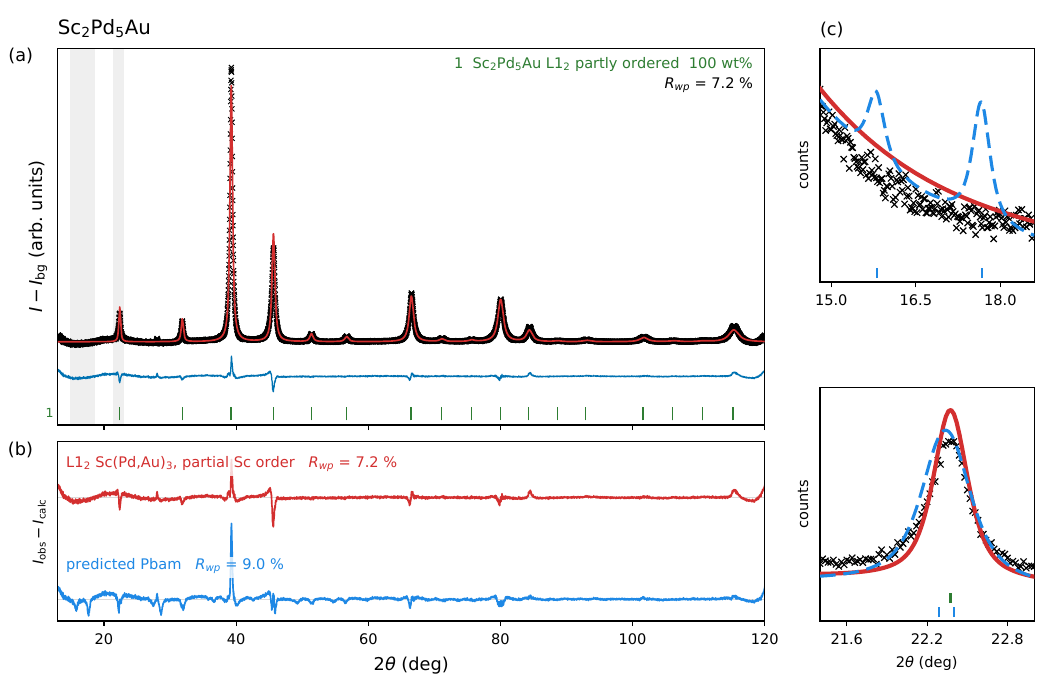}
\caption{As-cast \ch{Sc2Pd5Au}.
(b) L1$_2$ with Sc order and Pd and Au mixed (red) and the predicted $Pbam$ structure (blue).
(c) The $Pbam$ lines near \SI{15.7}{\degree} and \SI{17.6}{\degree} and the L1$_2$ (100) line at \SI{22.3}{\degree}.
Below \SI{17.5}{\degree} the fitted background runs above the data in both fits, so the absence of the $Pbam$ lines is judged from the shape of the data, not from the difference curve.}
\label{fig:S_row_16}
\end{figure}

\clearpage
\subsubsection{\ch{PtNiGe2}}\label{sec:S_row_17}

The predicted structure does not match the pattern of \ch{PtNiGe2}, but the fit is poor ($R_\mathrm{wp} = 21\%$), so this conclusion is provisional.
The predicted $P2_1/m$ structure orders Ni and Pt, and its strongest order line, (100) at \SI{15.9}{\degree}, is calculated at more than half the intensity of its strongest line.
We started both refinements at \SI{14}{\degree} to include this line, and the data there show a smooth tail with no peak.
A weaker order line at \SI{39.0}{\degree} is also absent.
The best match we found is MnP-type (Ni,Pt)Ge with Ni and Pt mixed, which at equal amounts of Ni and Pt has the nominal composition.
A strong line at \SI{32.2}{\degree}, about 70\% of the strongest observed line, is not explained by any model we tried.

\begin{figure}[!h]
\centering
\includegraphics[width=\textwidth]{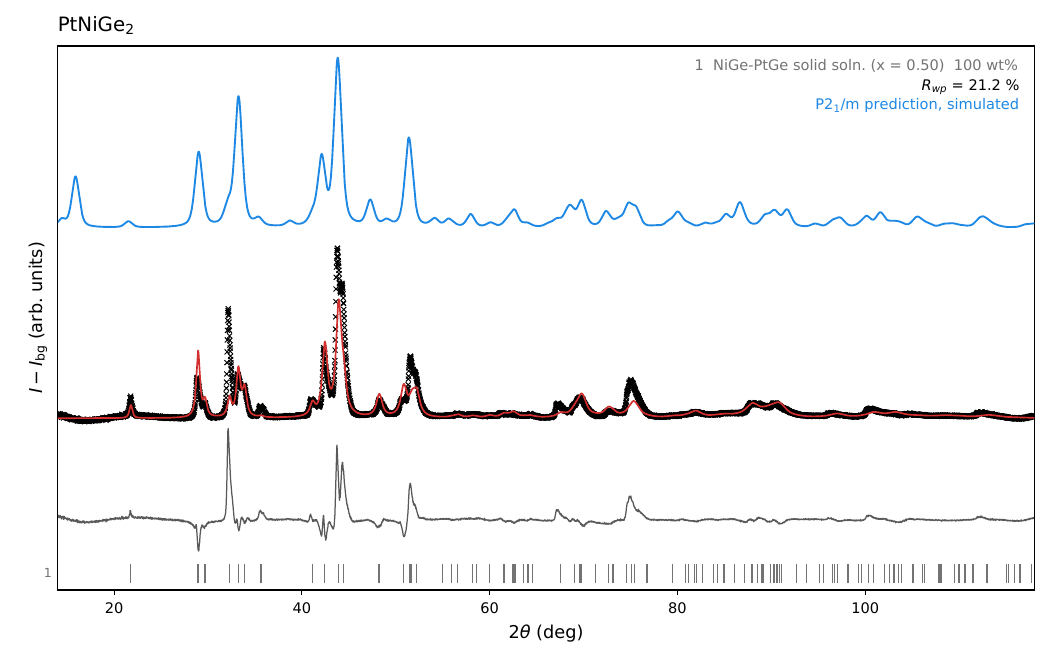}
\caption{\ch{PtNiGe2}. Measured pattern with the background subtracted (black crosses), the best fit (red) and its residual (gray), with the Bragg positions of its phases below. Above them, offset upward, the pattern of the predicted structure simulated from its computed crystal structure, with line widths taken from the measured pattern (blue), scaled to the strongest observed line.}
\label{fig:S_row_17}
\end{figure}

\clearpage
\subsubsection{\ch{Gd3Sn6Co}}\label{sec:S_row_18}

The predicted structure does not match the pattern of \ch{Gd3Sn6Co}.
The structure OQMD predicts is a $Cmcm$ structure with ordered Co, which shares its strong lines with the known CeNiSi$_2$-type GdCo$_x$Sn$_2$ and fits about as well overall.
The line that decides between them is the $Cmcm$ (152) line at \SI{36.5}{\degree}, which comes from the Co order.
This line is calculated at about forty times the noise, and the data show no line there (Fig.~\ref{fig:S_row_18}c).
The best match we found is GdCo$_x$Sn$_2$ alone, in which Co partly fills one site, refined with the structure of Gil et al.~\cite{Gil2004} ($R_\mathrm{wp} = 13.4\%$, Fig.~\ref{fig:S_row_18}).
So the predicted Co order did not form.

\begin{figure}[!h]
\centering
\includegraphics[width=\textwidth]{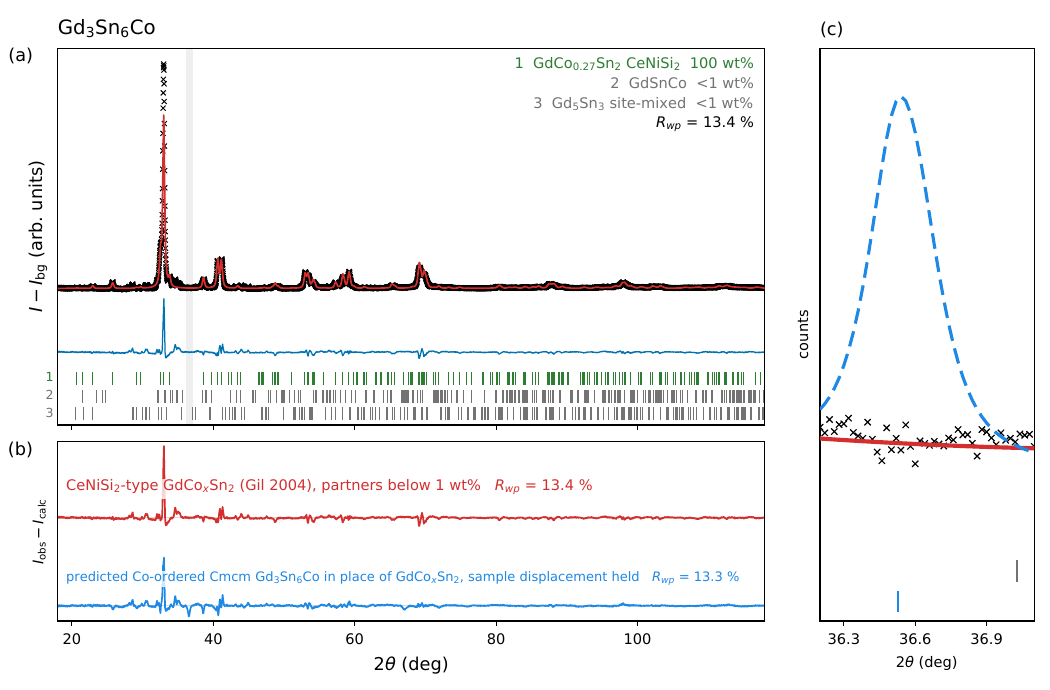}
\caption{As-cast \ch{Gd3Sn6Co}.
(b) CeNiSi$_2$-type GdCo$_x$Sn$_2$ (red) and the predicted Co-ordered $Cmcm$ structure in its place (blue).
(c) The (152) line of the prediction at \SI{36.5}{\degree}, absent from the data.}
\label{fig:S_row_18}
\end{figure}

\clearpage
\subsection{Samples that formed another solid solution}\label{sec:S_cat_othersolid}
\subsubsection{\ch{ScPd2Au}}\label{sec:S_row_19}

The lines of \ch{ScPd2Au} are broad, which hides much of the detail, and our best fit describes the pattern only approximately ($R_\mathrm{wp} = 7.9\%$, Fig.~\ref{fig:S_row_19}).
That fit uses two Sc(Pd,Au)$_3$ solid solutions that differ in their stacking, a hexagonal D0$_{24}$-type phase (about 58~wt\%) and a cubic L1$_2$ phase with Pd and Au mixed on the face centers.
Other combinations of disordered phases may describe the pattern as well.
The predicted $P4/mmm$ structure clearly does not fit.
With its DFT cell it gives $R_\mathrm{wp} = 18.6\%$, because its tetragonal cell splits the L1$_2$ (200) line into two lines, at 44.4 and \SI{46.5}{\degree}, where the data show a single line at \SI{45.5}{\degree}.
With the predicted Pd/Au order placed on the cubic cell, the superlattice line at \SI{31.6}{\degree} is calculated too strong ($R_\mathrm{wp} = 8.5\%$), so full Pd/Au order is unlikely.
A hexagonal line overlaps that position, so partial order cannot be excluded.
The predicted structure did not form, and the best fits come from disordered solid solutions.

\begin{figure}[!h]
\centering
\includegraphics[width=\textwidth]{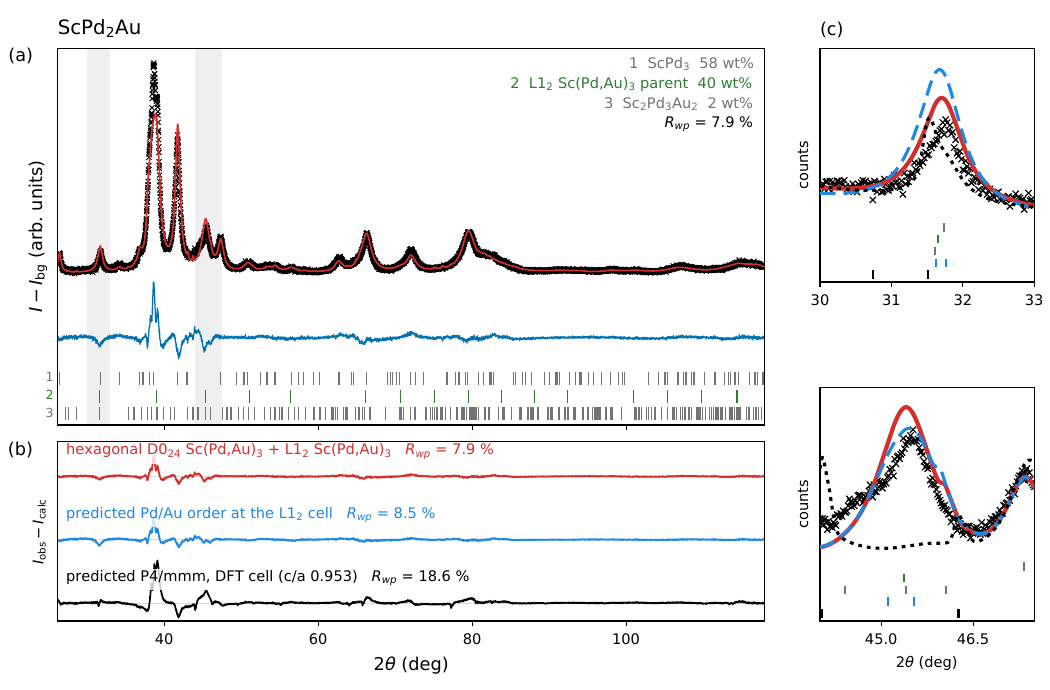}
\caption{\ch{ScPd2Au}.
(c) The superlattice line near \SI{31.6}{\degree} and the (200) line near \SI{45.5}{\degree}. Sample displacement held at zero (Sec.~\ref{sec:S_xrd}).}
\label{fig:S_row_19}
\end{figure}

\clearpage
\subsection{Samples that separated into several phases, none of them the on-hull structure}\label{sec:S_cat_multiphase}
\subsubsection{\ch{ScPd2Cu}}\label{sec:S_row_20}

The predicted D0$_{22}$ structure does not match the pattern of \ch{ScPd2Cu} well, and the fit we found is poor, so the description of what formed is provisional (Fig.~\ref{fig:S_row_20}).
Refined with lattice parameters near its DFT values, the D0$_{22}$ model fits much worse than our best match ($R_\mathrm{wp} = 12.8\%$ against 6.1\%).
Allowed to change its cell further, it becomes nearly cubic, so the data rule out the predicted cell shape but not the predicted Cu/Pd order.
The best match we found has two phases, an fcc-derived Sc(Cu,Pd)$_3$ phase with Cu and Pd mixed (about 61~wt\%) and a hexagonal \ch{CuPd3}-type phase.
A model with Cu and Pd ordered on the same fcc-derived lattice fits about as well ($R_\mathrm{wp} = 6.8\%$), because the lines are broad.
A sharp line at \SI{91.4}{\degree} is not explained by any model we tried.
The hexagonal model contains no Sc, so the composition of the fcc-derived phase is not known, and if it has the nominal composition it is a disordered version of the predicted structure.

\begin{figure}[!h]
\centering
\includegraphics[width=\textwidth]{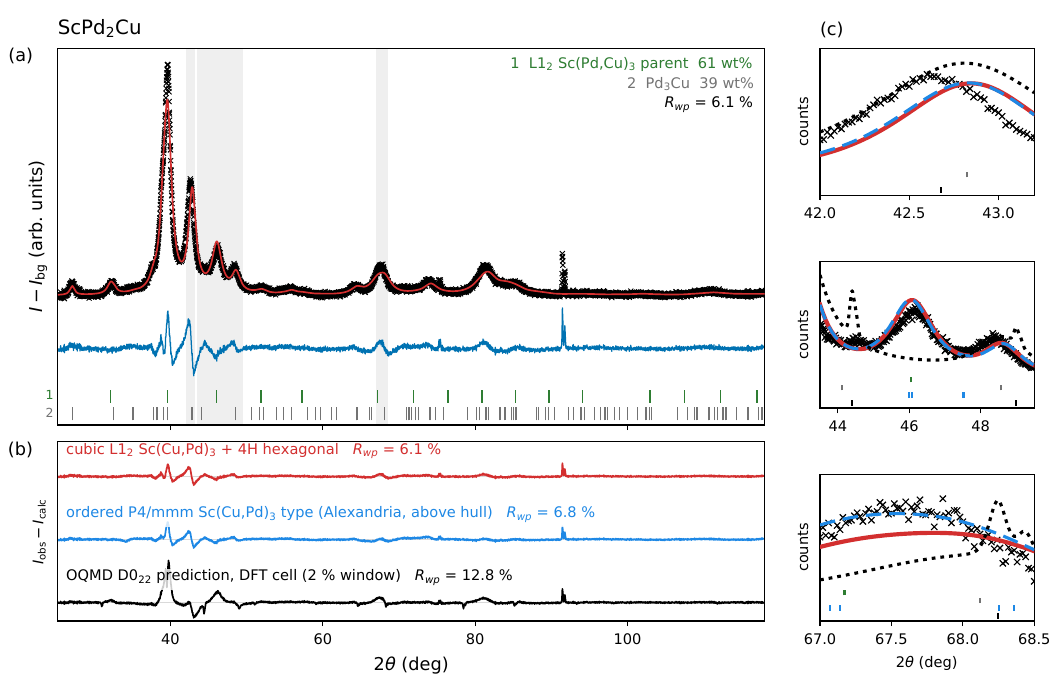}
\caption{\ch{ScPd2Cu}. (c) Lines near 42.6 and \SI{46.2}{\degree} and the (220) line. The lines are broad, and the disordered and ordered fcc-derived models differ little. Sample displacement held at zero (Sec.~\ref{sec:S_xrd}).}
\label{fig:S_row_20}
\end{figure}

\clearpage
\subsubsection{\ch{La5Al3Co2}}\label{sec:S_row_21}

The best match we found to the strongest lines of \ch{La5Al3Co2} is the known \ch{La4Co3Al3}, refined with the structure of Tursina et al.~\cite{Tursina2023} (Fig.~\ref{fig:S_row_21}).
Its presence is robust, because it accounts for the strongest lines.
It cannot be the whole sample, because its composition differs from the nominal one, and some lines are not explained by \ch{La4Co3Al3}, but we have not identified the other phases.
The predicted $Cmcm$ structure was not identified.

\begin{figure}[!h]
\centering
\includegraphics[width=\textwidth]{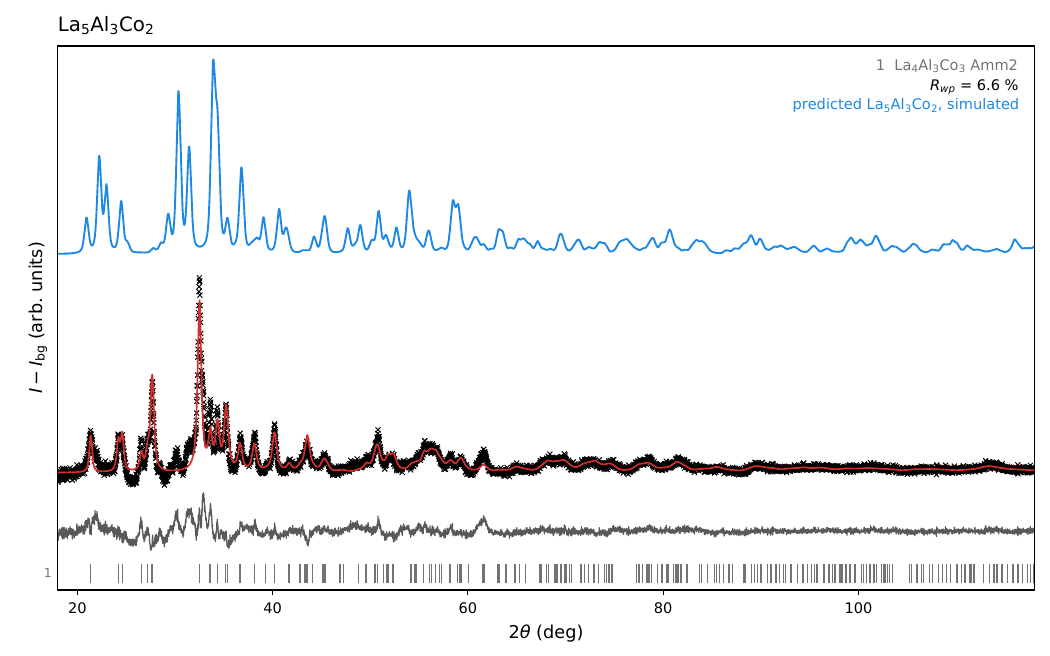}
\caption{\ch{La5Al3Co2}. Measured pattern with the background subtracted (black crosses), the best fit (red) and its residual (gray), with the Bragg positions of its phases below. Above them, offset upward, the pattern of the predicted structure simulated from its computed crystal structure, with line widths taken from the measured pattern (blue), scaled to the strongest observed line.}
\label{fig:S_row_21}
\end{figure}

\clearpage
\subsubsection{\ch{TiPt5Ge}}\label{sec:S_row_22}

The predicted structure does not match the pattern of \ch{TiPt5Ge}.
Refined alone, the predicted structure gives $R_\mathrm{wp} = 22.7\%$, and its line at \SI{31.66}{\degree}, clear of the Pt and \ch{Pt2Ge} lines, is absent from the data (Fig.~\ref{fig:S_row_22}).
A mixture of fcc Pt and Fe$_2$P-type \ch{Pt2Ge} matches the pattern well ($R_\mathrm{wp} = 5.4\%$).
Neither phase contains Ti, so a Ti-bearing phase, probably minor or poorly crystallized, is missing from the model.

\begin{figure}[!h]
\centering
\includegraphics[width=\textwidth]{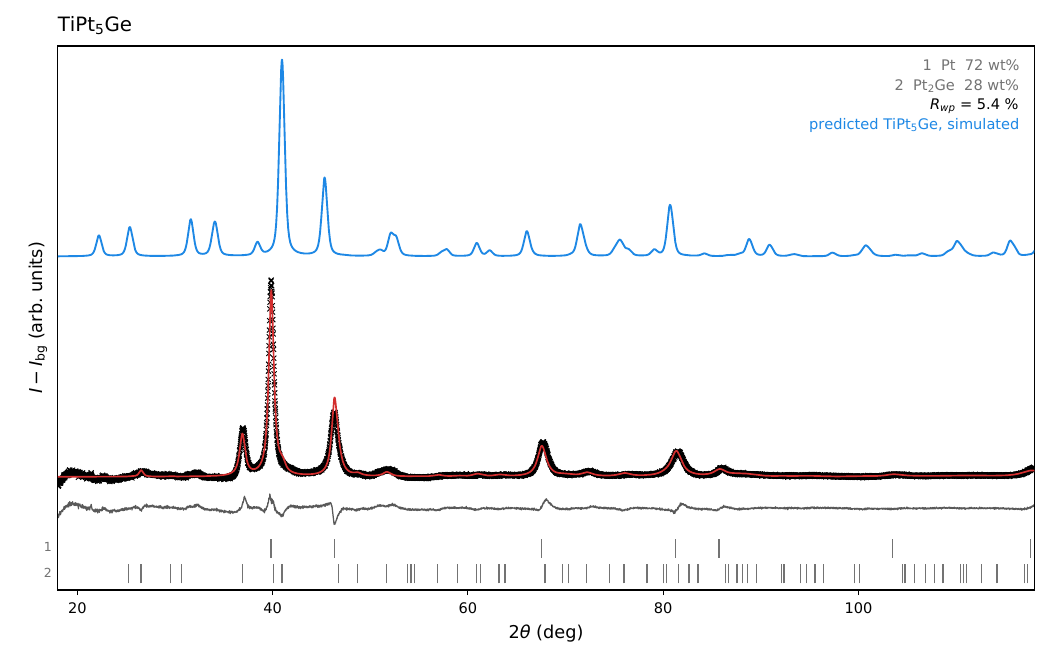}
\caption{\ch{TiPt5Ge}. Measured pattern with the background subtracted (black crosses), the best fit (red) and its residual (gray), with the Bragg positions of its phases below. Above them, offset upward, the pattern of the predicted structure simulated from its computed crystal structure, with line widths taken from the measured pattern (blue), scaled to the strongest observed line.}
\label{fig:S_row_22}
\end{figure}

\clearpage
\subsubsection{\ch{Ti2Pt8Si}}\label{sec:S_row_23}

The predicted structure does not match the pattern of \ch{Ti2Pt8Si}.
The best match we found has two phases in nearly equal amounts, a Pt-rich fcc-derived Ti--Pt phase, modeled with a \ch{TiPt2}-type cell, and a phase with the predicted $P4/mmm$ structure of \ch{TiPt5Si}.
The line near \SI{34.7}{\degree} belongs only to this \ch{TiPt5Si}-type phase (Fig.~\ref{fig:S_row_23}).
With the predicted \ch{Ti2Pt8Si} structure in its place, that line is left unexplained and $R_\mathrm{wp}$ rises from 7.9\% to 14.5\%.
The search thus found a predicted structure, that of \ch{TiPt5Si} (Sec.~\ref{sec:S_row_4}), in a sample whose nominal composition is different.

\begin{figure}[!h]
\centering
\includegraphics[width=\textwidth]{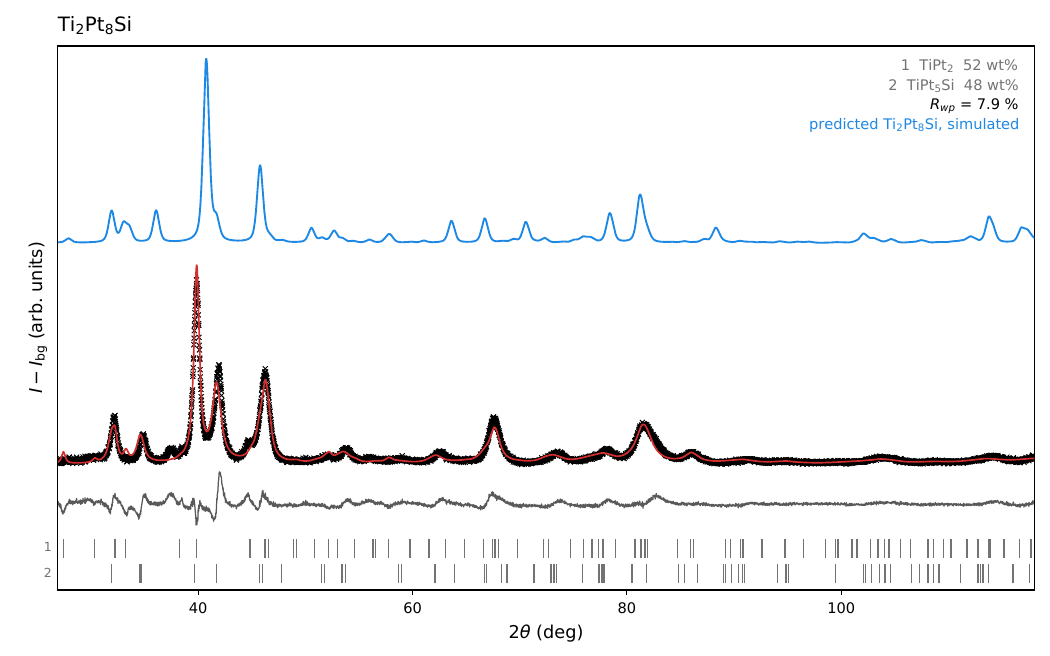}
\caption{\ch{Ti2Pt8Si}. Measured pattern with the background subtracted (black crosses), the best fit (red) and its residual (gray), with the Bragg positions of its phases below. Above them, offset upward, the pattern of the predicted structure simulated from its computed crystal structure, with line widths taken from the measured pattern (blue), scaled to the strongest observed line. Sample displacement held at zero (Sec.~\ref{sec:S_xrd}).}
\label{fig:S_row_23}
\end{figure}

\clearpage
\subsubsection{\ch{LaNiSi3}}\label{sec:S_row_24}

The predicted structure does not match the pattern of \ch{LaNiSi3}.
Refined alone, the predicted $I4mm$ structure gives $R_\mathrm{wp} = 41.3\%$ (Fig.~\ref{fig:S_row_24}).
A mixture of known La--Ni silicides, mainly \ch{La3Ni3Si7} and \ch{LaNiSi2}, with Si and probably some \ch{LaNi6Si6} in the structure of Pani et al.~\cite{Pani2014}, matches the pattern well ($R_\mathrm{wp} = 10.6\%$).

\begin{figure}[!h]
\centering
\includegraphics[width=\textwidth]{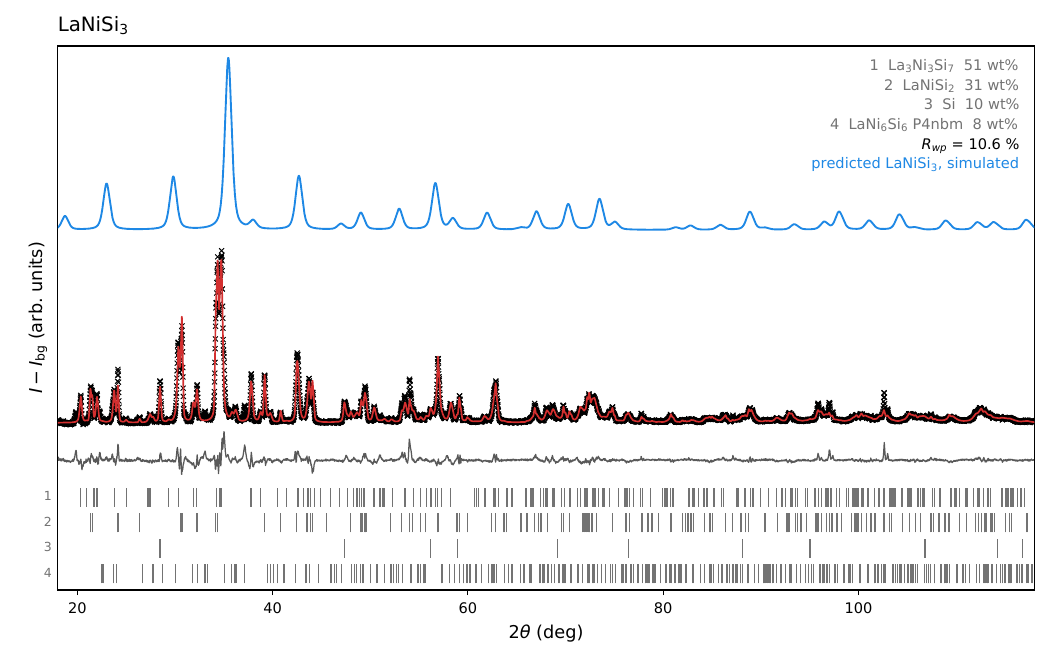}
\caption{\ch{LaNiSi3}. Measured pattern with the background subtracted (black crosses), the best fit (red) and its residual (gray), with the Bragg positions of its phases below. Above them, offset upward, the pattern of the predicted structure simulated from its computed crystal structure, with line widths taken from the measured pattern (blue), scaled to the strongest observed line.}
\label{fig:S_row_24}
\end{figure}

\clearpage
\subsubsection{\ch{GdSn6Co4}}\label{sec:S_row_25}

The predicted structure is not identified in the pattern of \ch{GdSn6Co4}.
Its strongest line coincides with lines of other phases, and at its next strongest lines, near 31.3, 41.7 and \SI{43.0}{\degree}, the data show no intensity beyond the fit.
The best match we found combines five known phases and fits the pattern well ($R_\mathrm{wp} = 7.8\%$, Fig.~\ref{fig:S_row_25}).
The two largest, a \ch{GdCo6Sn6}-type phase and \ch{Gd3Co4Sn13} (about 38 and 27~wt\%), have several lines of their own and are clearly present.
The three smaller ones, GdCo$_x$Sn$_2$ (refined with the structure of Gil et al.~\cite{Gil2004}), CoSn and Sn, are less certain.

\begin{figure}[!h]
\centering
\includegraphics[width=\textwidth]{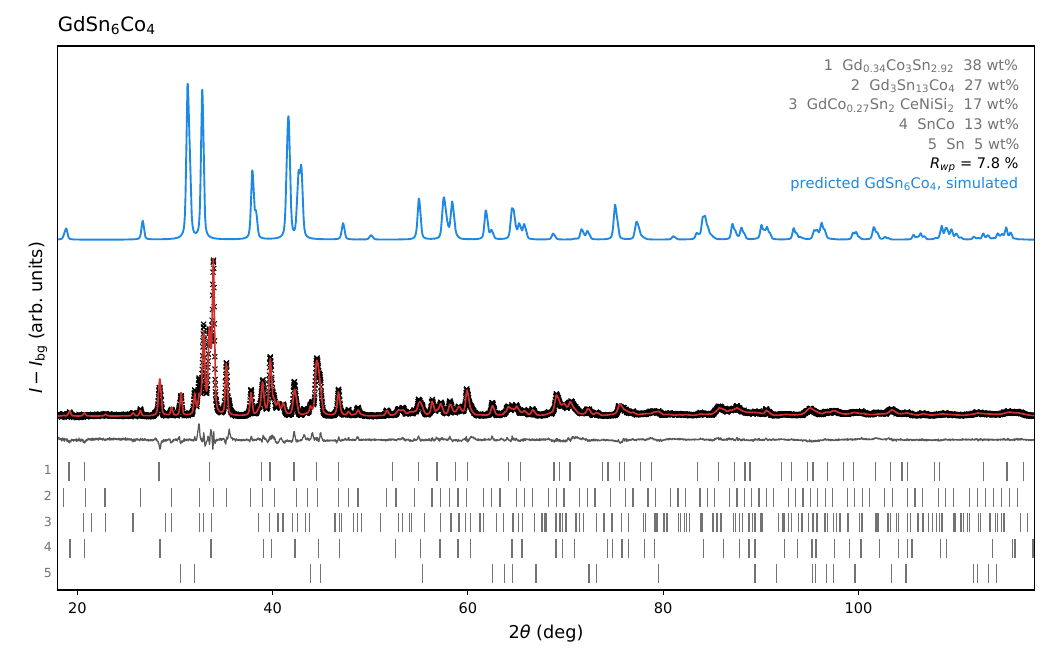}
\caption{\ch{GdSn6Co4}. Measured pattern with the background subtracted (black crosses), the best fit (red) and its residual (gray), with the Bragg positions of its phases below. Above them, offset upward, the pattern of the predicted structure simulated from its computed crystal structure, with line widths taken from the measured pattern (blue), scaled to the strongest observed line.}
\label{fig:S_row_25}
\end{figure}

\clearpage
\subsubsection{\ch{NbNi2B2}}\label{sec:S_row_26}

The predicted structure does not match the pattern of \ch{NbNi2B2}.
Refined alone, the predicted $I4/mmm$ structure gives $R_\mathrm{wp} = 38.4\%$, and its strong (101) line at \SI{28.40}{\degree} falls where the data show only background (Fig.~\ref{fig:S_row_26}).
A mixture of \ch{Ni3B}, \ch{NbB2} and the $\tau$-boride \ch{Nb2Ni21B6} matches the pattern well ($R_\mathrm{wp} = 11.3\%$).

\begin{figure}[!h]
\centering
\includegraphics[width=\textwidth]{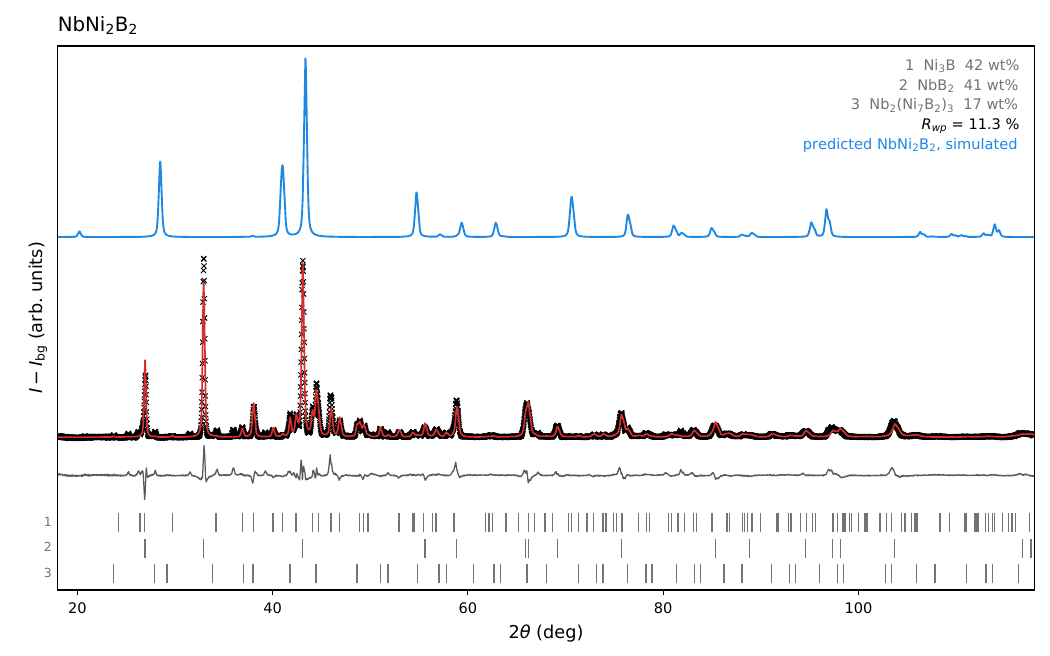}
\caption{\ch{NbNi2B2}. Measured pattern with the background subtracted (black crosses), the best fit (red) and its residual (gray), with the Bragg positions of its phases below. Above them, offset upward, the pattern of the predicted structure simulated from its computed crystal structure, with line widths taken from the measured pattern (blue), scaled to the strongest observed line.}
\label{fig:S_row_26}
\end{figure}

\clearpage
\subsubsection{\ch{GdFeGe2}}\label{sec:S_row_27}

The predicted structure does not match the pattern of \ch{GdFeGe2}.
Refined alone, the predicted $P2_1/m$ structure gives $R_\mathrm{wp} = 37.4\%$, and its isolated lines at 29.92 and \SI{42.91}{\degree} are absent (Fig.~\ref{fig:S_row_27}).
A mixture of known Gd--Fe--Ge phases, mainly \ch{GdFe_{0.37}Ge2} and \ch{GdFe2Ge2}, matches the pattern well ($R_\mathrm{wp} = 7.3\%$).

\begin{figure}[!h]
\centering
\includegraphics[width=\textwidth]{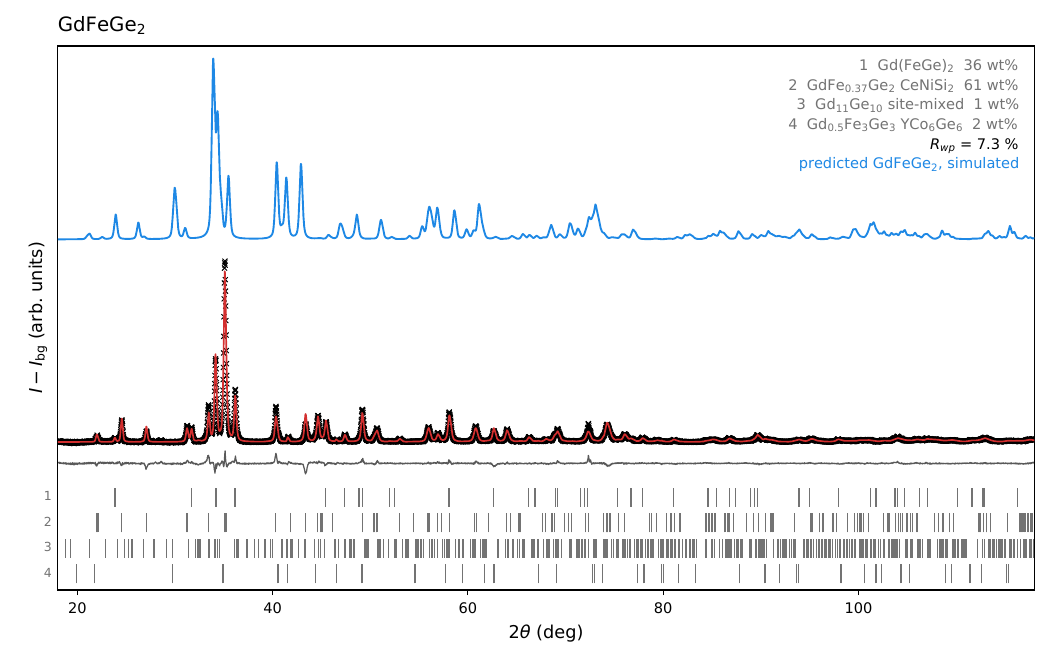}
\caption{\ch{GdFeGe2}. Measured pattern with the background subtracted (black crosses), the best fit (red) and its residual (gray), with the Bragg positions of its phases below. Above them, offset upward, the pattern of the predicted structure simulated from its computed crystal structure, with line widths taken from the measured pattern (blue), scaled to the strongest observed line.}
\label{fig:S_row_27}
\end{figure}

\clearpage
\subsubsection{\ch{Gd2Fe3Ge5}}\label{sec:S_row_28}

The predicted structure does not match the pattern of \ch{Gd2Fe3Ge5}.
Refined alone, the predicted $C2/c$ structure gives $R_\mathrm{wp} = 38.5\%$ (Fig.~\ref{fig:S_row_28}).
A mixture of known Gd--Fe--Ge phases, mainly \ch{GdFe_{0.37}Ge2}, \ch{GdFe2Ge2} and \ch{Gd_{0.5}Fe3Ge3}, with likely a little Ge, matches the pattern well ($R_\mathrm{wp} = 9.4\%$).

\begin{figure}[!h]
\centering
\includegraphics[width=\textwidth]{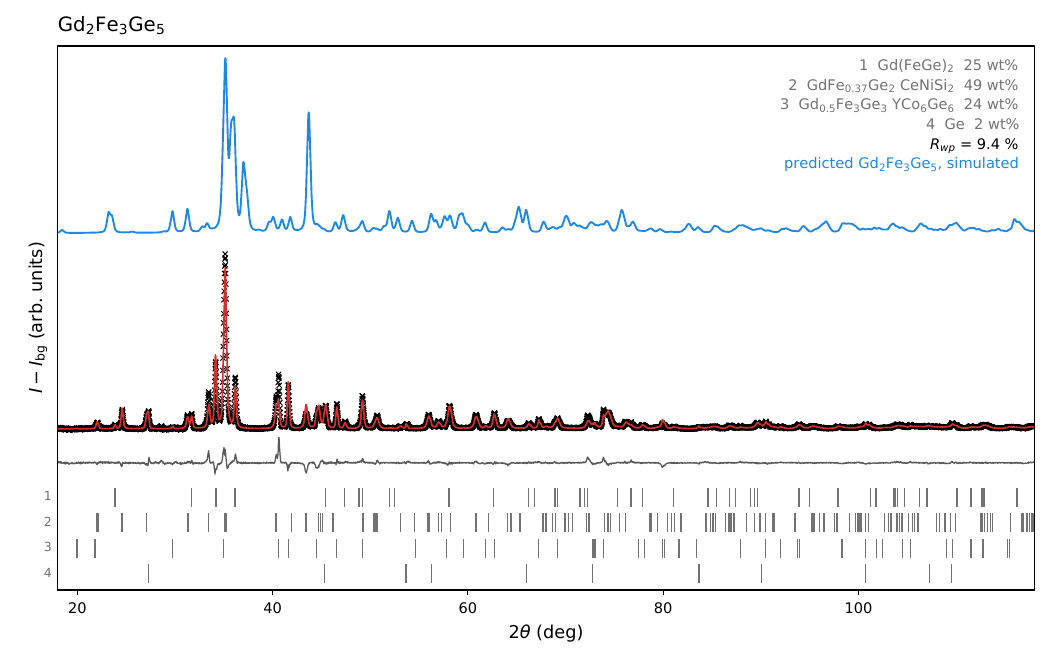}
\caption{\ch{Gd2Fe3Ge5}. Measured pattern with the background subtracted (black crosses), the best fit (red) and its residual (gray), with the Bragg positions of its phases below. Above them, offset upward, the pattern of the predicted structure simulated from its computed crystal structure, with line widths taken from the measured pattern (blue), scaled to the strongest observed line.}
\label{fig:S_row_28}
\end{figure}

\clearpage
\subsubsection{\ch{GdFeGe3}}\label{sec:S_row_29}

The predicted structure does not match the pattern of \ch{GdFeGe3}.
Refined alone, the predicted $I4mm$ structure gives $R_\mathrm{wp} = 39.9\%$, and its isolated (103) and (200) lines, at 34.43 and \SI{42.31}{\degree}, are absent (Fig.~\ref{fig:S_row_29}).
A mixture of \ch{GdFe_{0.37}Ge2}, \ch{Gd_{0.5}Fe3Ge3} and Ge matches the pattern well ($R_\mathrm{wp} = 7.5\%$).
A weak line at \SI{27.98}{\degree} is not explained.

\begin{figure}[!h]
\centering
\includegraphics[width=\textwidth]{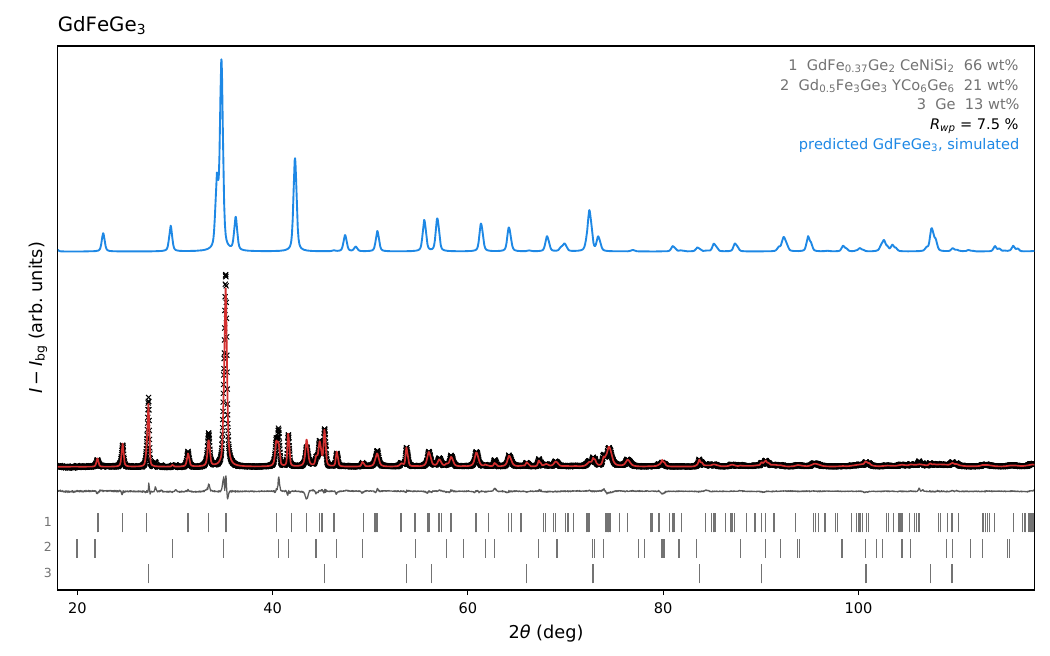}
\caption{\ch{GdFeGe3}. Measured pattern with the background subtracted (black crosses), the best fit (red) and its residual (gray), with the Bragg positions of its phases below. Above them, offset upward, the pattern of the predicted structure simulated from its computed crystal structure, with line widths taken from the measured pattern (blue), scaled to the strongest observed line.}
\label{fig:S_row_29}
\end{figure}

\clearpage
\subsubsection{\ch{LaRuC2}}\label{sec:S_row_30}

The predicted structure of \ch{LaRuC2} does not match the pattern of the button with the smallest mass loss (Fig.~\ref{fig:S_row_30}).
Refined alone it fits far worse than our best fit ($R_\mathrm{wp} = 40.6\%$ against 7.5\%).
The best fit uses Ru metal and La(OH)$_3$.
La(OH)$_3$ was not expected, and its presence suggests that the sample oxidized.
An earlier button, which lost more mass during melting, gave a different pattern that we could not describe, but the predicted structure did not form there either.

\begin{figure}[!h]
\centering
\includegraphics[width=\textwidth]{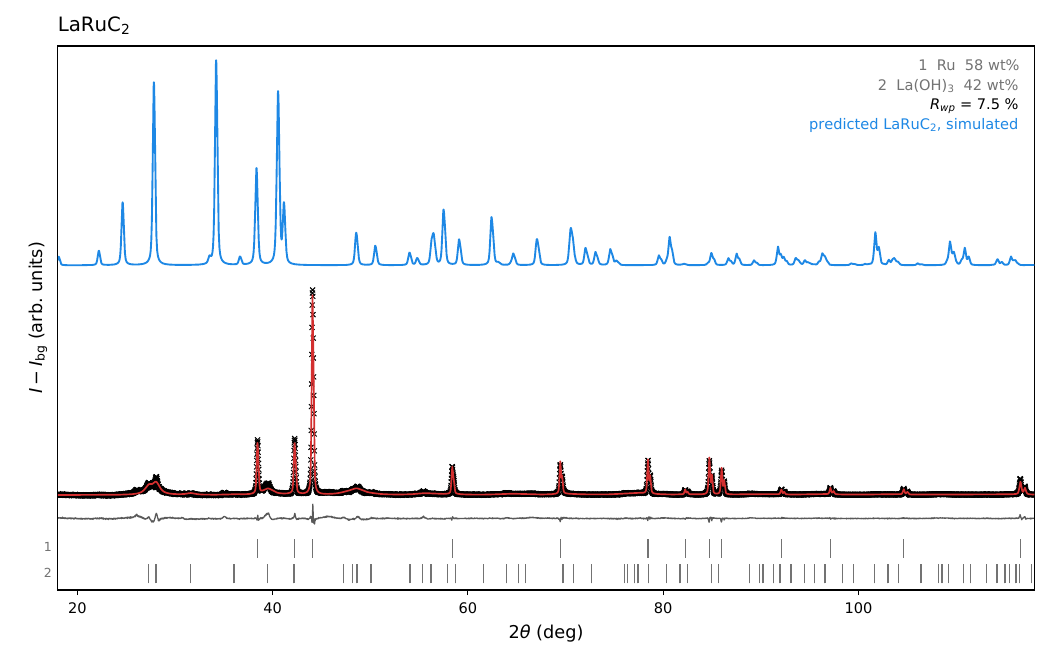}
\caption{\ch{LaRuC2}. Measured pattern with the background subtracted (black crosses), the best fit (red) and its residual (gray), with the Bragg positions of its phases below. Above them, offset upward, the pattern of the predicted structure simulated from its computed crystal structure, with line widths taken from the measured pattern (blue), scaled to the strongest observed line.}
\label{fig:S_row_30}
\end{figure}

\clearpage
\subsection{Samples whose phases are unresolved, without the on-hull structure}\label{sec:S_cat_unresolved}
\subsubsection{\ch{Pd6Cu5Ge}}\label{sec:S_row_31}

The predicted $Cmmm$ structure of \ch{Pd6Cu5Ge} does not match the pattern (Fig.~\ref{fig:S_row_31}).
Refined alone it fits much worse than our best fit ($R_\mathrm{wp} = 14.4\%$ against 8.7\%), and its superlattice lines near \SI{29.5}{\degree}, which would stand about six times the noise above the background, are absent.
The best fit uses a broad fcc solid solution and a small amount of a bcc phase.
Because the lines are so broad, this fit does not establish the composition or order of the fcc phase, so we quote no weight fractions and leave what formed unresolved.

\begin{figure}[!h]
\centering
\includegraphics[width=\textwidth]{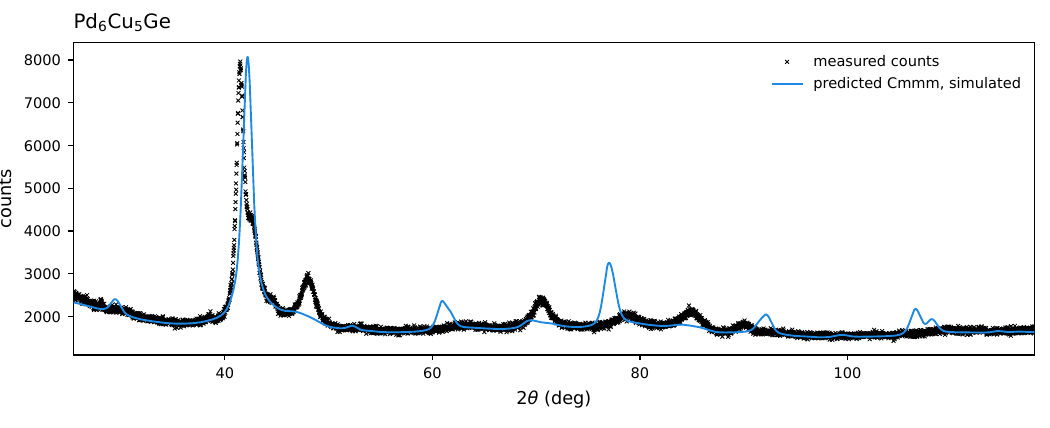}
\caption{\ch{Pd6Cu5Ge}, powder. Measured counts (black crosses) and the pattern of the predicted structure simulated from its computed crystal structure, with line widths taken from the measured pattern (blue), scaled to the strongest observed line and added to a smooth background estimate. No fit is shown because we have no satisfactory description of the pattern.}
\label{fig:S_row_31}
\end{figure}

\clearpage
\subsubsection{\ch{La2PdAu2}}\label{sec:S_row_32}

The pattern of \ch{La2PdAu2} is not well explained.
No combination of phases we tried fits it, and the best one misses about half the intensity of the strongest line, at \SI{32.54}{\degree} (Fig.~\ref{fig:S_row_32}).
The predicted $C2/m$ structure is not present, because its two strongest lines, at 31.13 and \SI{37.32}{\degree}, show no intensity above the background.
What formed is not identified.

\begin{figure}[!h]
\centering
\includegraphics[width=\textwidth]{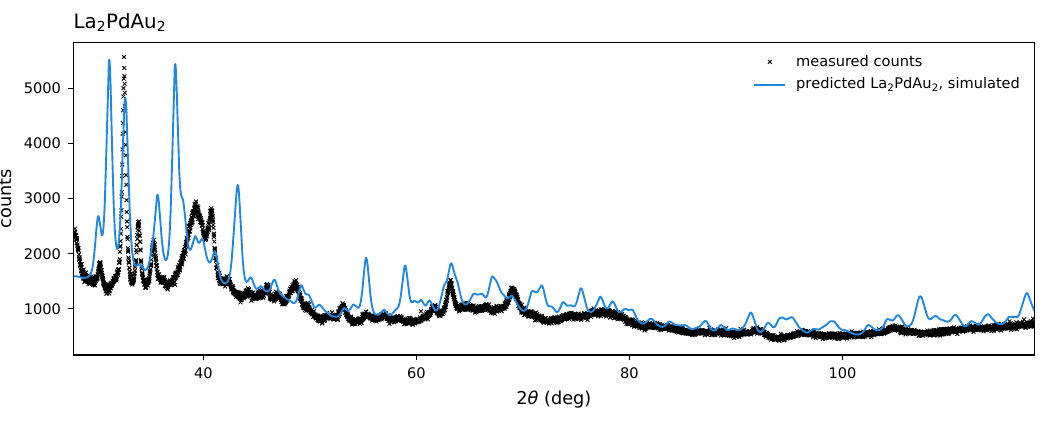}
\caption{\ch{La2PdAu2}. Measured counts (black crosses) and the pattern of the predicted structure simulated from its computed crystal structure, with line widths taken from the measured pattern (blue), scaled to the strongest observed line and added to a smooth background estimate. No fit is shown because we have no satisfactory description of the pattern.}
\label{fig:S_row_32}
\end{figure}

\clearpage
\subsubsection{\ch{Ti2Pt8Ge}}\label{sec:S_row_33}

The predicted $P4/mmm$ structure of \ch{Ti2Pt8Ge} does not match the pattern, which was measured on a saw-cut face of the as-cast button (Fig.~\ref{fig:S_row_33}).
The strongest observed lines index on a nearly cubic, Pt-rich fcc-type cell, while the tetragonal cell of the prediction places lines at 40.4 and \SI{49.8}{\degree}, where the data show none.
We have no satisfactory description of the whole pattern, so what formed is not identified.

\begin{figure}[!h]
\centering
\includegraphics[width=\textwidth]{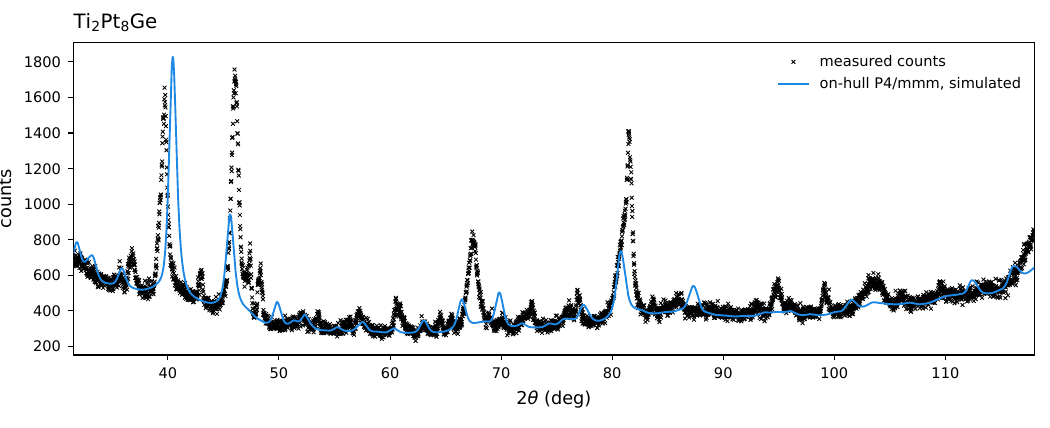}
\caption{\ch{Ti2Pt8Ge}, as-cast, measured on a saw-cut face. Measured counts (black crosses) and the pattern of the predicted structure simulated from its computed crystal structure, with line widths taken from the measured pattern (blue), scaled to the strongest observed line and added to a smooth background estimate. No fit is shown because we have no satisfactory description of the pattern.}
\label{fig:S_row_33}
\end{figure}

\clearpage
\subsubsection{\ch{ScPdCu4}}\label{sec:S_row_34}

The patterns of \ch{ScPdCu4} show only broad features, without well-defined lines (Fig.~\ref{fig:S_row_34}).
The three buttons we measured gave similar patterns (Table~\ref{tab:S_buttons}).
The pattern computed from the predicted structure does not match them, and we could not identify the phases that formed.

\begin{figure}[!h]
\centering
\includegraphics[width=\textwidth]{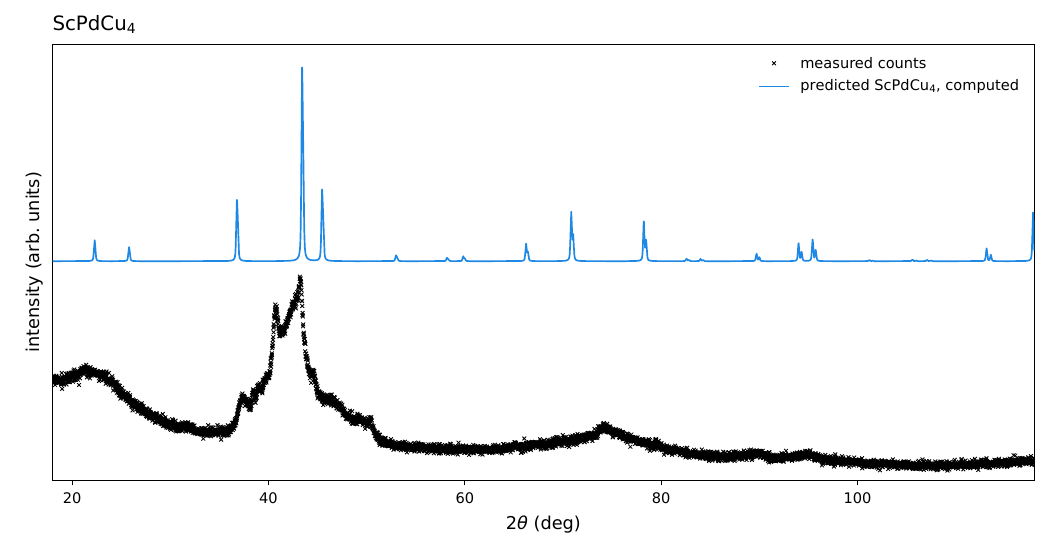}
\caption{\ch{ScPdCu4}. Measured counts (black crosses) and, above them, the pattern computed from the predicted structure with a line width of \SI{0.15}{\degree} and no background (blue). No fit is shown because we have no satisfactory description of the pattern.}
\label{fig:S_row_34}
\end{figure}

\fi

\end{document}